\documentclass[aps,prb,10pt,amsmath,amssymb,footinbib,showpacs,twocolumn]{revtex4-2}
\usepackage{amsmath}
\usepackage{amssymb}
\usepackage{amsthm}
\usepackage{setspace}
\usepackage{graphicx}
\usepackage{braket}
\usepackage{mathrsfs}
\usepackage{float}
\usepackage[colorlinks = true,linkcolor = blue,urlcolor  = blue,citecolor = blue,anchorcolor = blue]{hyperref}
\usepackage[utf8]{inputenc}
\usepackage[english]{babel}
\usepackage{bm}

\begin{document}

\title{Geometry, anomaly, topology, and transport in Weyl fermions}
\author{Azaz Ahmad}
\affiliation{School of Physical Sciences, Indian Institute of Technology Mandi, Mandi 175005, India.}
\author{Gautham Varma K.}
\affiliation{School of Physical Sciences, Indian Institute of Technology Mandi, Mandi 175005, India.}
\author{Gargee Sharma}
\affiliation{School of Physical Sciences, Indian Institute of Technology Mandi, Mandi 175005, India.}

\begin{abstract}
Weyl fermions are one of the simplest objects that link ideas in geometry and topology to high-energy physics and condensed matter physics. Although the existence of Weyl fermions as elementary particles remains dubious, there is mounting evidence of their existence as quasiparticles in certain condensed matter systems. Such systems are termed Weyl semimetals (WSMs). Needless to say, WSMs have emerged as a fascinating class of materials with unique electronic properties, offering a rich playground for both fundamental research and potential technological applications. This review examines recent advancements in understanding electron transport in Weyl semimetals (WSMs). We begin with a pedagogical introduction to the geometric and topological concepts critical to understanding quantum transport in Weyl fermions. We then explore chiral anomaly (CA), a defining feature of WSMs, and its impact on transport phenomena such as longitudinal magnetoconductance (LMC) and the planar Hall effect (PHE). The Maxwell-Boltzmann transport theory extended beyond the standard relaxation-time approximation is then discussed in the context of  Weyl fermions, which is used to evaluate various transport properties. Attention is also given to the effects of strain-induced gauge fields and external magnetic fields in both time-reversal broken and inversion asymmetric inhomogeneous WSMs. The review synthesizes theoretical insights, experimental observations, and numerical simulations to provide a comprehensive understanding of the complex transport behaviors in WSMs, aiming to bridge the gap between theoretical predictions and experimental verification.
\end{abstract}

\maketitle
\tableofcontents
\section{Introduction} 
\subsection{Overview}
The role of topology in quantum condensed matter systems has been a subject of extensive study and exploration for the past forty years. The genesis can be traced back to the discovery of the integer and fractional quantum Hall effects in the 1980s~\cite{klitzing1980new,tsui1982two}, which were the first few examples of states of matter topologically distinct from all previously known states. For example, the integer quantum Hall state of matter does not break any symmetries but is characterized by a topological invariant, which leads to 
a precise quantization of the Hall conductance ($n e^2/h$) regardless of the microscopic details~\cite{laughlin1981quantized,thouless1982quantized}. The fractional quantum Hall states lead to fractional quantization of Hall conductance [$(p/q)e^2/h$] arising from an interplay between interactions and topology~\cite{laughlin1983anomalous}. 
The study of the quantum Hall effects and its ramifications led to a new classification paradigm centered on the concept of topological order~\cite{wen1995topological}. Concurrently, ideas establishing the role of geometry in quantum mechanics have flourished as well ever since Berry explored {holonomy} in adiabatic electronic transport~\cite{berry1984quantal}. The study of topology and geometry in condensed matter gained rapid attention in the 2000s following the discovery of graphene and topological insulators~\cite{hasan2010colloquium,qi2011topological,neto2009electronic,sarma2011electronic}. 
Today, the roles of topology, geometry, and the Berry phase in electronic transport in crystals are now firmly established~\cite{haldane2017nobel,xiao2010berry,wen2017colloquium,volovik2003universe}. The understanding and manipulation of these concepts have opened up new avenues for research and potential applications in electronic devices and quantum computing.

In a seemingly unrelated context, the history of Weyl and Dirac fermions spans nearly a hundred years. In 1928, Dirac~\cite{dirac1928quantum} attempted the quantum solution of a relativistic electron and derived the eponymous Dirac equation, which describes an electron of mass $m$ and momentum $\textbf{p}$ as a four-component spinor dispersing as $\epsilon_\mathbf{p} = \sqrt{p^2c^2+m^2c^4}$. When $m=0$, Dirac's solution can be reformulated as two distinct two-component fermions with opposite chiralities, known as Weyl fermions~\cite{weyl1929gravitation}. Since the past ten years, Weyl fermions (WFs) have surprisingly resurged in condensed matter physics by appearing as quasiparticle excitations in a class of certain (semi)metallic systems, known as Weyl semimetals (WSMs)~\cite{hosur2013recent,armitage2018weyl}. It turns out that a WSM phase can appear at the critical point between a topological and a trivial insulator, serving as an intermediate phase in the topological phase transition~\cite{murakami2007phase,murakami2007tuning,xu2011chern}. WSM is itself a stable and topological phase, characterized by gapless Weyl fermionic excitations in the bulk, which are protected by translation symmetry. Furthermore, they exhibit peculiar Fermi arc surface states, which are projections of the gapless points in the Brillouin zone~\cite{wan2011topological,burkov2011topological,burkov2011weyl}. Weyl fermions thus lie at the intersection of topology, geometry, high-energy physics, and condensed matter, making its study highly rewarding from multiple perspectives.
Over the past decade, several theoretical predictions and experimental verifications of the WSM phase have been made in systems such as TaAs, NbAs, TaP, NbP, MoTe2, and WTe2~\cite{xu2015discovery,xu2016observation,shekhar2015extremely,soluyanov2015type,lv2015observation,yang2015weyl,zhang2016linear,weng2015weyl,arnold2016chiral,klotz2016quantum,liang2015ultrahigh,huang2015weyl,xu2016spin,hasan2021weyl}.

The flow of electrons in WSMs is a problem that has garnered significant attention as well. What makes electron transport fascinating in WSMs is the unique interplay between their geometric and topological properties and high-energy physics phenomena, enabling the emergence of exotic, anomalous, and topological features not found in conventional metallic compounds. The most famous effect in WFs is the so-called chiral anomaly (CA). It originates from high-energy physics~\cite{adler1969axial,bell1969pcac}, where the conservation of left- and right-handed Weyl fermions is violated in the presence of non-orthogonal electric and magnetic fields. This anomaly has reemerged in the study of WSMs and has attracted significant interest in the condensed-matter community~\cite{armitage2018weyl,volovik2003universe,nielsen1981no,nielsen1983adler,wan2011topological,xu2011chern,zyuzin2012weyl,son2013chiral,goswami2013axionic,goswami2015axial,zhong2015optical,kim2014boltzmann,lundgren2014thermoelectric,cortijo2016linear,sharma2016nernst,zyuzin2017magnetotransport,das2019berry,kundu2020magnetotransport,knoll2020negative,sharma2020sign,bednik2020magnetotransport,he2014quantum,liang2015ultrahigh,zhang2016signatures,li2016chiral,xiong2015evidence,hirschberger2016chiral}. In WSMs, which host Weyl fermions as quasiparticle excitations, chiral anomaly is also expected to manifest under external electromagnetic fields. Key transport signatures of this anomaly include positive longitudinal magnetoconductance (LMC)~\cite{son2013chiral} and the planar Hall effect (PHE)~\cite{nandy2017chiral}. 
Intense efforts have been devoted to understanding these anomaly-induced conductivities in WSMs~\cite{zyuzin2012weyl,son2013chiral,goswami2013axionic,goswami2015axial,zhong2015optical,kim2014boltzmann,lundgren2014thermoelectric,cortijo2016linear,sharma2016nernst,zyuzin2017magnetotransport,das2019berry,kundu2020magnetotransport,knoll2020negative,sharma2020sign,bednik2020magnetotransport,das2019linear,sharma2019transverse,sharma2023decoupling,ahmad2021longitudinal,ahmad2023longitudinal,varma2024magnetotransport,sharma2017chiral,sharma2017nernst,nandy2017chiral,he2014quantum,zhang2016signatures,li2016chiral,xiong2015evidence,hirschberger2016chiral}. Additionally, non-electronic probes such as optical processes can also indicate the presence of a chiral anomaly~\cite{goswami2015optical,levy2020optical,parent2020magneto,song2016detecting,rinkel2017signatures,yuan2020discovery,cheng2019probing}. Remarkably, strain induces axial vector fields in Weyl semimetals and affects its electronic and thermal transport, adding on to the list of unconventional behavior realized in WSMs~\cite{grushin2016inhomogeneous,ahmad2023longitudinal}.

In this work, we present a comprehensive overview of some recent developments in electronic transport studies of WSMs. In Sec.~\ref{sec:geom}, we start by presenting an intuitive picture of geometry, curvature, and topology in the context of quantum transport. In Sec.~\ref{sec:qg and wf} and \ref{sec:wf and cond matt}, we then connect the seemingly different areas of quantum geometry and WFs, and their relevance in modern condensed matter physics. Sec.~\ref{sec:ca and WF} introduces chiral anomaly in Weyl fermions. Section~\ref{sec2} presents the Boltzmann electronic transport formalism for Weyl semimetals, while Sec.~\ref{sec:sec3} presents recent results of longitudinal magnetoconductance, planar Hall effect, and the effects of strain in inhomogeneous WSMs. 

\begin{figure}
    \centering
    \includegraphics[width=\columnwidth]{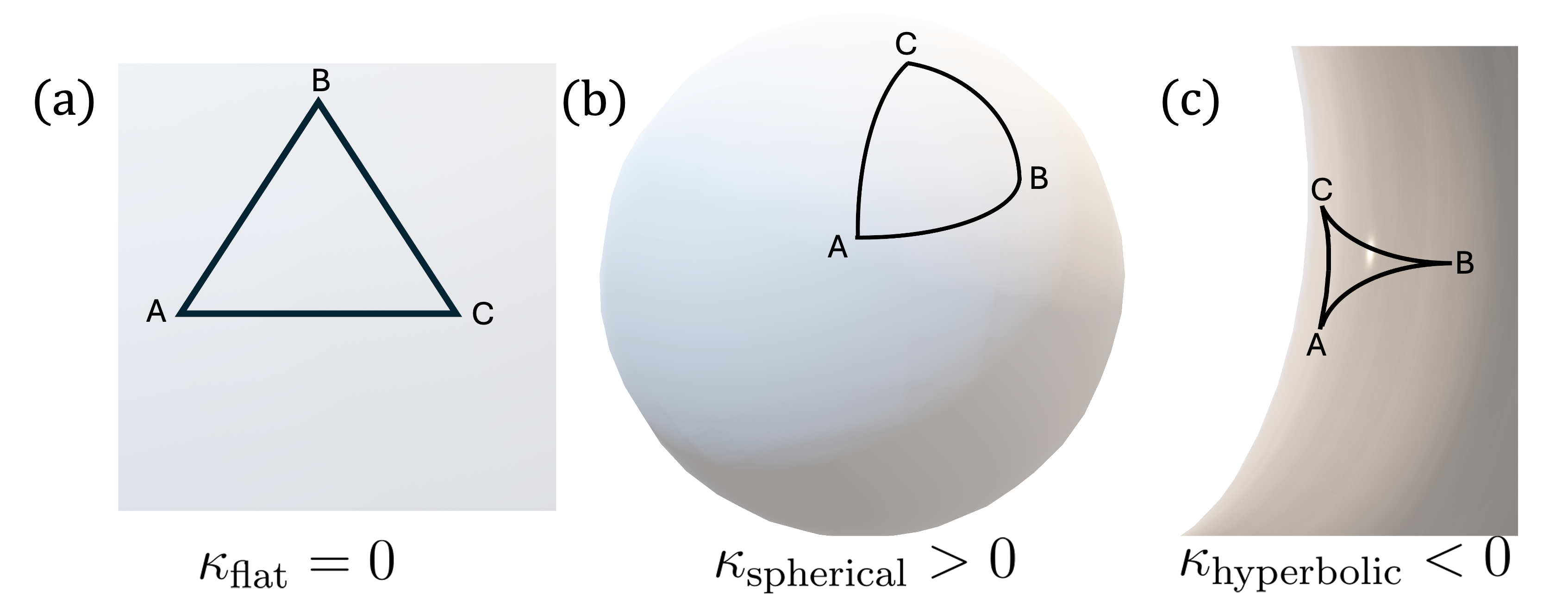}
    \caption{Geometry of different surfaces can be characterized by their curvature $\kappa$. The properties of objects existing on surfaces of different curvatures are different. For example, the sum of angles of a triangle drawn equals (a) $\pi$ for a flat surface, (b) greater than $\pi$ for a spherical, and (c) less than $\pi$ for a hyperbolic surface.}
    \label{fig_triangles}
\end{figure}

\subsection{Geometry and quantum physics}
\label{sec:geom}
The geometry of space is fundamental in determining the properties of objects existing in it. The following example makes this more apparent. Consider a triangle in a flat space. Euclidean geometry predicts that the sum of angles of a triangle equals $\pi$. If one instead attempts to draw a triangle on a sphere, straight lines in flat space become geodesics, which connect two points by the shortest distance. In this case, the sum of the angles of the triangle is always greater than $\pi$, which is a characteristic of spherical geometry. If we repeat the exercise on a surface with hyperbolic geometry, the sum of the angles of a triangle is now less than $\pi$ (see Fig.~\ref{fig_triangles}). One can then define curvature ($\kappa$) as the angular excess $\epsilon(\triangle)$ (sum of the angles of a triangle minus $\pi$) per unit area $\mathcal{A}(\triangle)$~\cite{needham2021visual}: 
\begin{align}
    \kappa = \frac{\Sigma_{\mathrm{angles}} (\triangle)-\pi}{\mathrm{area}} \equiv \frac{\epsilon(\triangle)}{\mathcal{A}(\triangle)}.
\end{align}
The geometry of the three spaces discussed above can then be characterized by their respective curvatures (also see Fig.~\ref{fig_triangles}): 
\begin{align}
    \kappa_\mathrm{flat}&=0\nonumber\\
    \kappa_\mathrm{spherical}&>0\nonumber\\
    \kappa_\mathrm{hyperbolic}&<0\nonumber.
\end{align}
On more general surfaces the curvature may not be constant, and one can instead define a local (or Gaussian) curvature $\kappa_p$ at every point $p$ on the surface:
\begin{align}
    \kappa_p = \lim_{\triangle_p\rightarrow p}\frac{\epsilon(\triangle_p)}{\mathcal{A}(\triangle_p)}.
\end{align}
For example, the inner surface of a torus has $\kappa<0$ but the outer surface has $\kappa>0$. 

\begin{figure}
    \centering
    \includegraphics[width=\columnwidth]{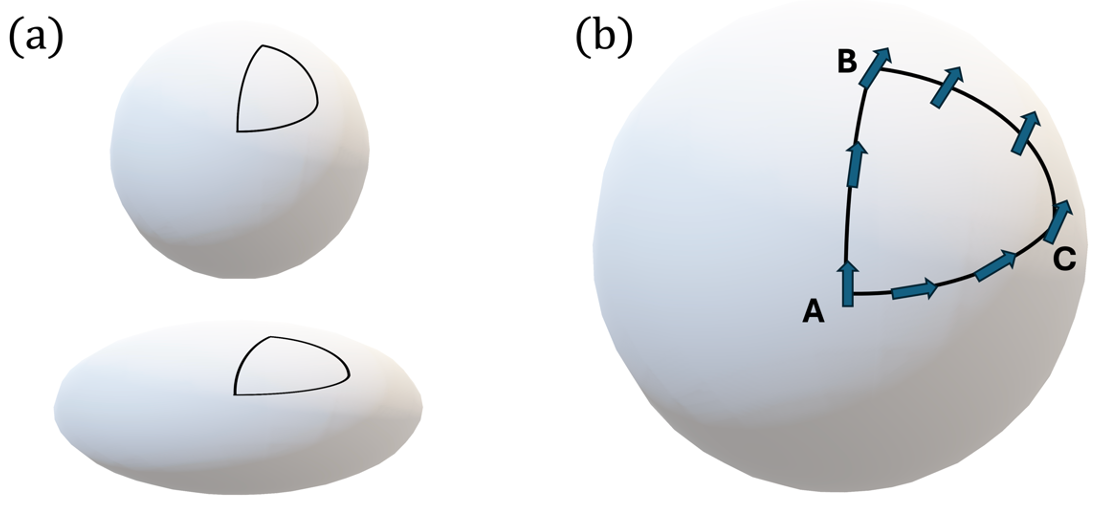}
    \caption{(a) Homeomorphism stretches a sphere to a spheroid but preserves the total curvature. (b) Parallel transporting a vector on the surface of a sphere along the path $A\rightarrow B\rightarrow C\rightarrow A$ results in an angular offset (holonomy).}
    \label{fig_parallel}
\end{figure}

We can then imagine smoothly stretching a surface such that the lengths and angles of a triangle drawn on the surface are not preserved but no additional cuts or holes are introduced or removed from the surface (see Fig.~\ref{fig_parallel} (a)). Mathematically, such a transformation is termed \textit{homeomorphism}. A famous theorem by Gauss (the Gauss-Bonnet theorem) states that the total curvature of the surface is conserved under {homeomorphism}~\cite{needham2021visual, nakahara2018geometry}:
\begin{align}
\Sigma_p\kappa(p)\xrightarrow[]{\mathrm{homeomorphism}}\mathrm{constant}.
\end{align}
The Gauss-Bonnet theorem provides a deep and remarkable connection between geometry and topology and has influenced several branches of mathematics and physics. We will see this shortly, but for now we imagine parallel transporting a vector on the spherical surface and move it along the geodesic triangle (see Fig.~\ref{fig_parallel} (b)). It is easy to verify that the vector rotates when it returns to its starting point. The net rotation (mathematically termed as $\textit{holonomy}$) is equal to the angular excess of the geodesic triangle and is also equal to the total curvature of the triangle.
\begin{align}
\mathrm{holonomy}=\mathrm{angular}\hspace{1mm}{\mathrm{excess}}=\mathrm{total} \hspace{1mm}{\mathrm{curvature}}.
\end{align}
Although we discuss this for a triangular path in our example, a non-trivial holonomy may arise for any closed path. In other words, a vector $v=(a, b)$ acquires a phase (implemented by a rotation $\mathcal{R}(\theta)$) after parallel transport in a closed loop on a surface, i.e., 
\begin{align}
    v=\begin{pmatrix}
        a\\
        b
    \end{pmatrix}\xrightarrow[|a|^2+|b|^2=|a'|^2+|b'|^2]{\mathrm{parallel}\hspace{0.5mm} \mathrm{transport}}\begin{pmatrix}
        a'\\
        b'
    \end{pmatrix}=\mathcal{R}(\theta) v.
\end{align}

In quantum mechanics, the state of a system can be represented by a vector in Hilbert space. It is therefore natural to consider the effect of parallel transporting a quantum state $|n\rangle$ adiabatically in a closed loop on a curved surface. For example, this could be implemented by suitably varying a parameter $R_i$ of the underlying Hamiltonian $H(\textbf{R}(t))$ as the state evolves in time. The quantum state $|n\rangle$ rotates or in other words, acquires a phase [that can be referred as $\exp(i\gamma)$] at the end of the adiabatic evolution, where $\gamma$ is given by~\cite{berry1984quantal} 
\begin{align}
    \gamma = i\oint_c \langle n | \nabla n\rangle\cdot \mathbf{R},
\end{align}
where the gradient is taken with respect to the parameter $\mathbf{R}$. This is also known as the Berry phase~\cite{berry1984quantal}, which can also be expressed as an integral of the Gaussian curvature ($\boldsymbol{\Omega}(\mathbf{R})$) of the surface enclosed inside the closed loop:
\begin{align}
    \gamma = \int\boldsymbol{\Omega}(\mathbf{R})\cdot d\mathbf{S},
\end{align}
where $\boldsymbol{\Omega}(\mathbf{R})=i\nabla\times \langle n | \nabla n\rangle$ is also known as the Berry curvature. It turns out that the Berry phase is gauge invariant and is fully consistent with time-dependent quantum evolution. Since $\gamma$ is invariant under homeomorphism, it is also a topological invariant. 
\subsection{Quantum geometry and Weyl fermions}
\label{sec:qg and wf}
To elucidate the non-trivial role of the Berry phase, we consider a simple two-band Hamiltonian of the form $H_\mathbf{k} = \mathbf{k}\cdot\boldsymbol{\sigma}$, where $\mathbf{k}$ is a vector in three-dimensions, and $\boldsymbol{\sigma}$ represents the vector of Pauli matrices.
\begin{align}
    H_\mathbf{k} = \mathbf{k}\cdot\boldsymbol{\sigma} = \begin{pmatrix}
        k_z & k_x - i k_y \\
        k_x + i k_y & -k_z
    \end{pmatrix}.
    \label{Eq_H_weyl}
\end{align}
In general, $\boldsymbol{\sigma}$ represents any suitable degree of freedom: orbital, spin, or sublattice, and $\mathbf{k}$  represents any suitable physical parameter. 
\begin{figure}
    \centering
    \includegraphics[width=\columnwidth]{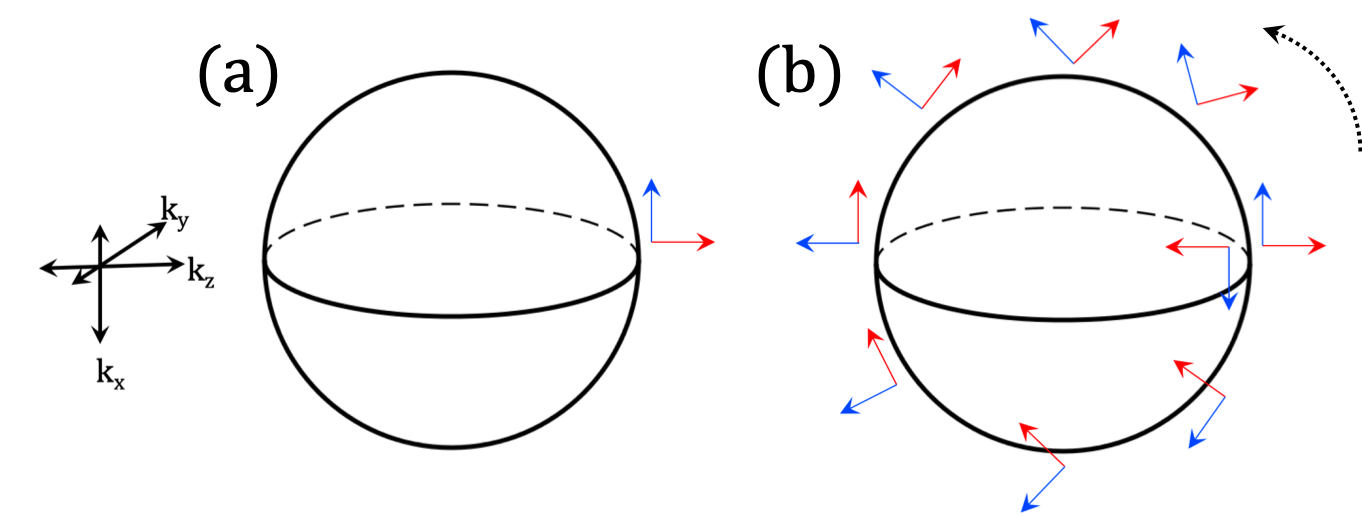}
    \caption{(a) Eigenstates of the Hamiltonian $H_\mathbf{k}$ (eigenframe) in Eq.~\ref{Eq_H_weyl} represented by the orthogonal red and blue arrows. (b) Evolving the eigenframe in the direction of the dotted arrow by varying the parameter $\theta$ from zero to $2\pi$ rotates it by $\pi$.}
    \label{fig_sphere_weyl}
\end{figure}
The eigenenergies of the above Hamiltonian are $E_\mathbf{k}=\pm k$ and the eigenstates are $|n\rangle^+ = (\cos(\theta/2) e^{-i\phi}, \sin(\theta/2))^\mathrm{T}$, $|n\rangle^- = (\sin(\theta/2) e^{-i\phi}, -\cos(\theta/2))^\mathrm{T}$, where $\theta$ is the polar angle $(\theta=\cos^{-1}k_z/k)$, and $\phi$ is the azimuthal angle $(\tan\phi=k_y/k_x)$. We evolve the eigenstates in a fixed gauge varying the polar angle $\theta = \cos^{-1}k_z/k$ (see Fig.~\ref{fig_sphere_weyl}). As $\theta$ varies from zero to $2\pi$, the eigenstates acquire a nontrivial $\pi$ phase. This is illustrated by the $\pi$ eigenframe rotation in Fig.~\ref{fig_sphere_weyl} (b).

The Berry curvature for the above Hamiltonian is easily evaluated to be
\begin{align}
    \boldsymbol{\Omega}_\mathbf{k}^{\pm} = \mp \frac{\hat{k}}{2 k^2}.
\end{align}
We note that this is the vector field with a singular point at the origin. This scenario is similar to the electric field generated by a point change, and thus we can say that the Berry curvature is generated by a monopole at the degeneracy point $\mathbf{k}=0$~\cite{dirac1931quantised,wu1975concept,xiao2010berry}. The degeneracy point thus acts as a source or a sink of the Berry curvature. Integrating the Berry curvature around the degeneracy point equals the number of monopoles (in this case one) in units of $2\pi$. As long as the degeneracy points are protected, homeomorphism on the Hamiltonian by employing small perturbations conserves the number of monopoles or in other words conserves the total curvature by Gauss-Bonnet theorem. This idea lies at the heart of topological protection of quantum states. 

As an illustrative comparison, we consider the following Hamiltonian 
\begin{align}
    H_\mathbf{k}=\begin{pmatrix}
        k_z^2-k_x^2-k_y^2 & 2k_z \sqrt{k_x^2+k_y^2}\\
        2k_z \sqrt{k_x^2+k_y^2} & k_x^2+k_y^2-k_z^2
    \end{pmatrix},
\end{align}
which also has the same energy spectrum $E_\mathbf{k}=\pm k$, but the eigenframe in this case rotates by $2\pi$ as $\theta$ rotates by $2\pi$, entailing a trivial quantum geometry compared to Eq.~\ref{Eq_H_weyl}. 

Surprisingly, the Hamiltonian in Eq.~\ref{Eq_H_weyl} is identical to the Hamiltonian of Weyl fermions originating in high-energy physics~\cite{weyl1929gravitation}. Weyl fermions here refer to the massless solutions of the Dirac equation~\cite{dirac1928quantum}. When the mass term in the Dirac equation is set to zero, the Dirac Hamiltonian can be expressed in a block-diagonal form, also known as the Weyl Hamiltonian: 
\begin{align}
        H_\mathbf{k} = \begin{pmatrix}
            \mathbf{k}\cdot\boldsymbol{\sigma} & 0 \\
            0 & -\mathbf{k}\cdot\boldsymbol{\sigma}
        \end{pmatrix}
\end{align}.
The diagonal entries in the above Hamiltonian represent Weyl fermions of opposite chiralities (plus/minus or equivalently left/right)~\cite{peskin2018introduction}. 
In elementary particle physics neutrinos were initially thought to be Weyl fermions but it was later discovered that neutrinos have a finite mass. So far Weyl fermions have not been observed to exist as elementary particles. 
\begin{figure*}
    \centering
    \includegraphics[width=2\columnwidth]{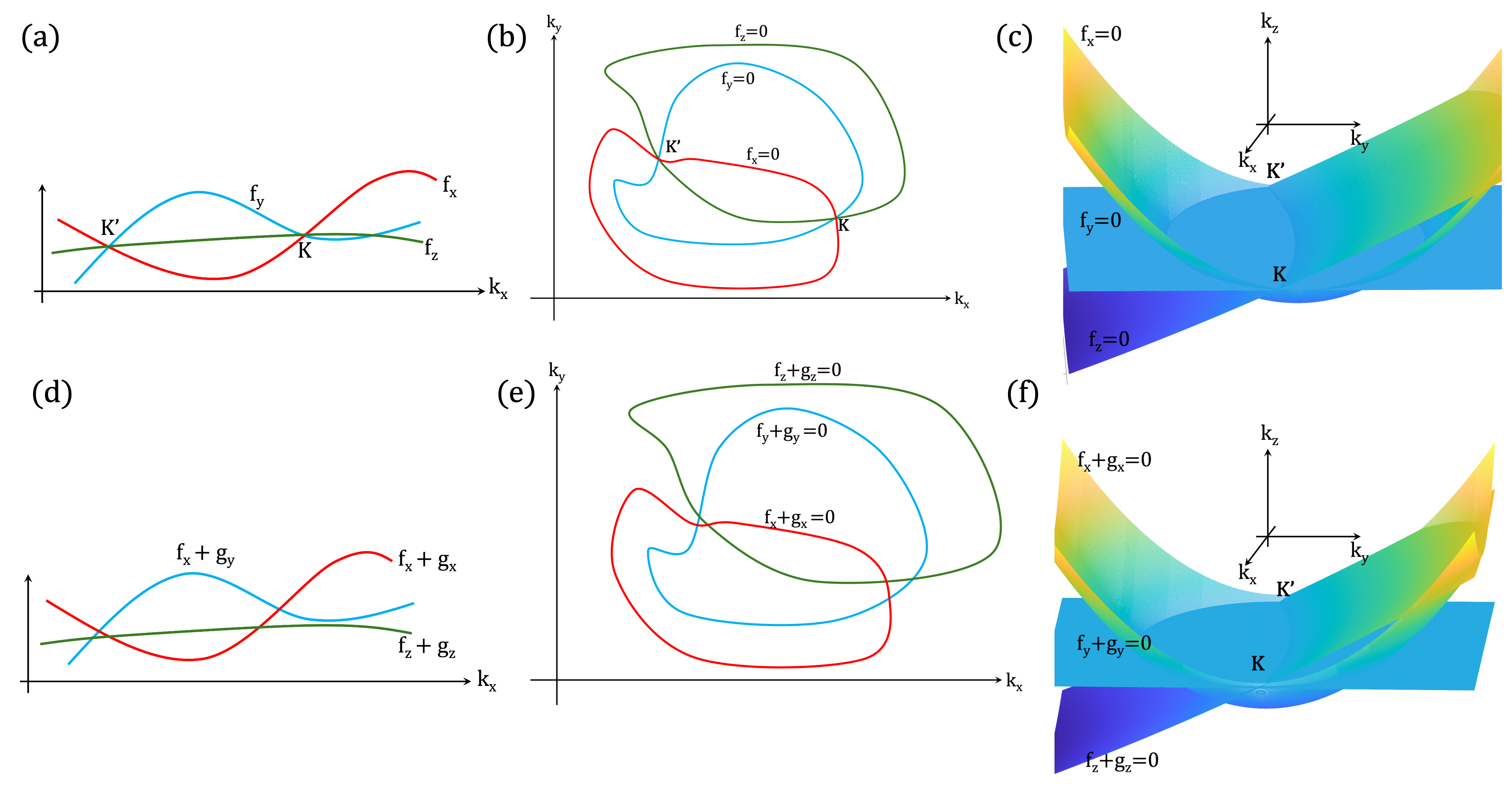}
    \caption{(a) Degenerate points in $d=1$ correspond to three functions intersecting at two points $K$ and $K'$. (b) When $d=2$, this condition is modified to three intersecting curves on a plane. (c) When $d=3$, the condition modifies to three intersecting surfaces meeting at two points $\mathbf{K}$ and $\mathbf{K}$'. (d) and (e) Small perturbations ($\mathbf{g}_\mathbf{k}$) move the curves and the fine-tuned Weyl points are destroyed. (f) Small perturbations move the surfaces and the Weyl points shift in momentum-space but remain protected.}
    \label{fig_bandtouch}
\end{figure*}
\subsection{Topological Weyl fermions in condensed matter physics}
\label{sec:wf and cond matt}
The occurrence of degenerate points in Bloch bands of crystals was first examined by Herring~\cite{herring1937accidental} who argued that band degeneracies could occur accidentally in solids. This raised the possibility of observing Weyl fermions as quasiparticle excitations around the degeneracy point where the Berry curvature in the periodic momentum space (Brillouin zone) can become singular. 
However, it was much later that Nielsen and Ninomiya studied chiral Weyl fermions on a lattice and pointed out that they must occur in pairs~\cite{nielsen1981no, nielsen1983adler}. The proof relies on the observation that every source of the Berry curvature vector field must have a sink so that the sum of the indices of the vector field on the Brillouin zone equals zero~\cite{smit2003introduction, friedan1982proof}. 

The occurrence and topological protection of Weyl fermions as quasiparticles in three spatial dimensions can be understood by considering a general two-band Hamiltonian of the form:~\cite{armitage2018weyl} 
\begin{align}
    H_\mathbf{k} = \boldsymbol{\sigma}\cdot\mathbf{f}_\mathbf{k} = \begin{pmatrix}
        f_z(\mathbf{k}) & f_x(\mathbf{k}) - i f_y(\mathbf{k})\\
        f_x(\mathbf{k}) + i f_y(\mathbf{k}) & -f_z(\mathbf{k})
    \end{pmatrix}.
    \label{Eq_H_f_weyl}
\end{align}
In the above Hamiltonian $\mathbf{f}_\mathbf{k}$ is a general vector-valued function of crystal-momentum $\mathbf{k}$. The Hamiltonian can describe the low-energy band structure comprising two bands lying close to the Fermi energy in any spatial dimension $d$. The energy spectrum is given by $\epsilon_\mathbf{k} = \pm \sqrt{f_x(\mathbf{k})^2 + f_y(\mathbf{k})^2 + f_z(\mathbf{k})^2}$. Note that we measure energy relative to the midgap or the band degeneracy point. An energy shift can be easily accomplished by adding a term $\epsilon_0\mathbb{I}_{2\times 2}$ in the Hamiltonian. The condition for band-degeneracy is given by $f_x(\mathbf{k})^2 + f_y(\mathbf{k})^2 + f_z(\mathbf{k})^2=0$. Let us assume that when $d=1$,  band-degeneracy occurs at a points ${K}$ and $K'$, i.e., $f_x(K)=f_y(K)=f_z(K)=0$, and $f_x(K')=f_y(K')=f_z(K')=0$. This requires three scalar functions of a single variable to intersect at points $K$ and $K'$. This condition may be satisfied with some amount of fine-tuning (Fig.~\ref{fig_bandtouch} (a)). We now add a perturbative disorder to the Hamiltonian of the form $V_\mathbf{k} = \boldsymbol{\sigma}\cdot \mathbf{g}_\mathbf{k}$. The degeneracy condition at point $K$ is then modified to $f_x(K)+g_x(K)=f_y(K)+g_y(K)=f_z(K)+g_z(K)=0$, and a similar condition exists for the point $K'$ (Fig.~\ref{fig_bandtouch} (d)). In general, this is very hard to satisfy because of the random nature of the disorder. The band degeneracy points are therefore not protected and the degeneracy can be lifted by infinitesimal disorder. 

In two spatial dimensions ($d=2$), the band-degeneracy condition at points $\mathbf{K}=({K_x,K_y})$ and $\mathbf{K}'=({K'_x,K'_y})$ is $f_x(\mathbf{K})=f_y(\mathbf{K})=f_z(\mathbf{K})=0$, and $f_x(\mathbf{K}')=f_y(\mathbf{K}')=f_z(\mathbf{K}')=0$. This requires three curves to intersect precisely at two points, which again may be realized by fine-tuning (Fig.~\ref{fig_bandtouch} (b)). Small perturbations can move the curves (Fig.~\ref{fig_bandtouch} (e)) on the plane and thus the degenerate point is again not protected. Moving to $d=3$, the band-degeneracy condition at point $\mathbf{K}=({K_x,K_y,K_z})$ and $\mathbf{K}=({K_x',K_y',K_z'})$ is $f_x(\mathbf{K})=f_y(\mathbf{K})=f_z(\mathbf{K})=0$, and a similar condition for $\mathbf{K}'$. This condition amounts to three intersecting surfaces meeting at two points (Fig.~\ref{fig_bandtouch} (c)). Perturbations will change the surfaces and modify the degeneracy conditions (Fig.~\ref{fig_bandtouch} (f)). However, the degenerate points are just shifted in momentum space, unless the two points move closer and annihilate. Band touching is therefore robust in three spatial dimensions. 

\begin{figure}
    \centering
    \includegraphics[width=.95\columnwidth]
    {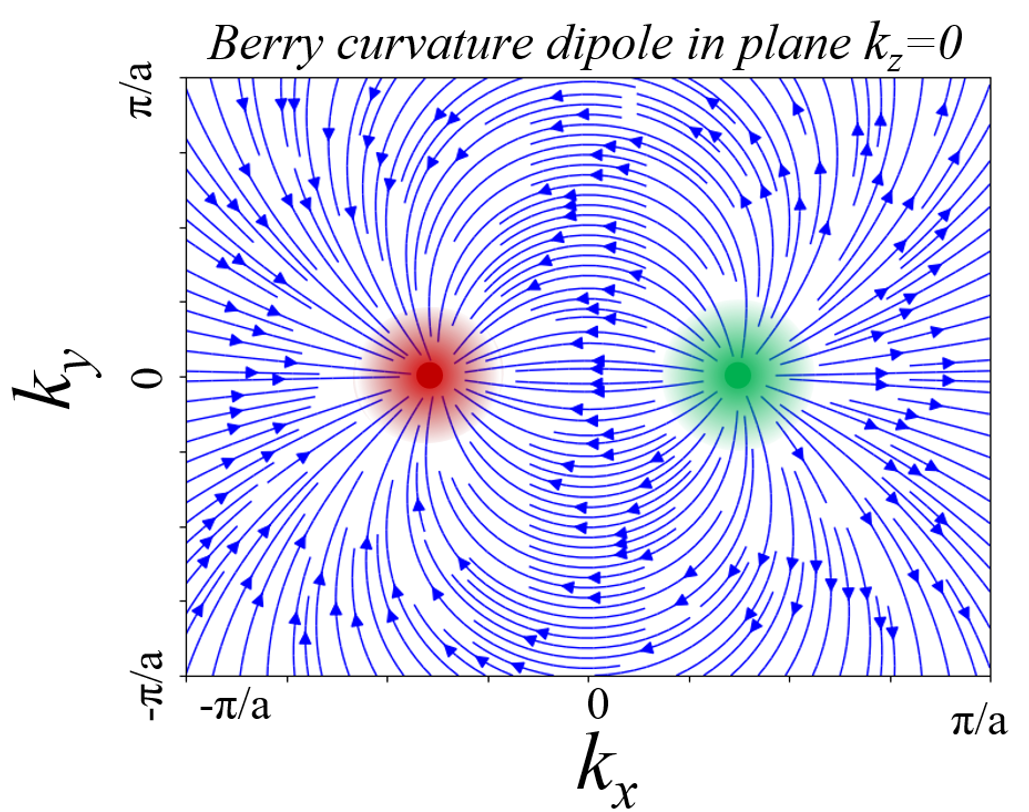}
    \caption{Schematic diagram to show the Berry curvature field lines in momentum space for $k_z=0$. The red and green regions correspond to two points having opposite chirality. Here, $a$ is the lattice constant of the system.}
\label{fig:BC_dipole}
\end{figure}

Having found protected degenerate points in a three-dimensional solid, we can now expand the Hamiltonian in Eq.~\ref{Eq_H_f_weyl} around the point $\mathbf{K}$ as: 
\begin{align}
    H_\mathbf{k} = \begin{pmatrix}
        \nabla f_z(\mathbf{K}) &\nabla f_x(\mathbf{K})-i\nabla f_y(\mathbf{K})\\ 
        \nabla f_x(\mathbf{K})+i\nabla f_y(\mathbf{K})&-\nabla f_z(\mathbf{K})
    \end{pmatrix}\cdot\mathbf{k},
\end{align}
where $\mathbf{k}$ is now measured relative to the $\mathbf{K}$ point. The above Hamiltonian can be expressed as:
\begin{align}
    H_\mathbf{k} = \sum_{i=x,y,z}\sum_{j=x,y,z}v_{ij}k_i \sigma_j,
\end{align}
where $v_{ij}=\partial_i f_{j}$. This is the general form of the Weyl fermion Hamiltonian near the Weyl node. The nodal point is protected by the \textit{chirality} quantum number $\chi$, given by $\chi = \mathrm{sign} (\mathrm{det} [v_{ij}])$, which is also equal to the flux of the Berry curvature of a Bloch band: 
\begin{align}
    \chi = \frac{1}{2\pi}\oint{\boldsymbol{\Omega}_\mathbf{k}\cdot d\mathbf{S}}.
\end{align}
Here the integral is over the Fermi surface enclosing the nodal point and $d\mathbf{S}$ is the area element.
In an appropriate frame of reference we may write $v_{ij}=v_{i}\delta_{ij}$, and obtain a more familiar form of the Weyl Hamiltonian $H_{\mathbf{k}}=\Tilde{{\mathbf{k}}}\cdot\boldsymbol{\sigma}$, where $\Tilde{\mathbf{k}}={(v_xk_x, v_yk_y, v_zk_z)}$ similar to Eq.~\ref{Eq_H_weyl}. A similar expansion can be done around the $\mathbf{K}'$ point. Although $H_{\mathbf{k}}=\Tilde{{\mathbf{k}}}\cdot\boldsymbol{\sigma}$ is anisotropic, much of the fundamental Weyl physics can be understood by considering the isotropic version $H_\mathbf{k}=\mathbf{k}\cdot\boldsymbol{\sigma}$. The Berry curvature of a pair of Weyl nodes is schematically depicted in Fig.~\ref{fig:BC_dipole}.
\subsection{Chiral anomaly of Weyl fermions}
\label{sec:ca and WF}
Chiral anomaly (CA) refers to the non-conservation of the left and right-handed chiral Weyl fermions in the presence of external gauge fields. It is also known as the Adler-Bell-Jackiw (ABJ) anomaly and originates in high energy physics \cite{adler1969axial}. This non-conservation of chiral charges results in a nonvanishing chiral current that may lead to chirality-dependent transport. In the context of WSMs, CA can be verified experimentally through the measurement of (but not limited to) magnetoconductance~\cite{son2013chiral}, Hall conductance~\cite{burkov2011weyl,burkov2014anomalous}, thermoelectric~\cite{lundgren2014thermoelectric} and Nernst effects~\cite{sharma2016nernst}, optical processes~\cite{goswami2015optical}, and non-local transport~\cite{parameswaran2014probing}. 

Nielson and Ninomiya were among the first to investigate chiral anomaly in crystals~\cite{nielsen1981no,nielsen1983adler}. They derived the chiral anomaly or the ABJ anomaly from a physical point of view as the production of Weyl particles, and show that there is an absence of the net production of particles for local chiral invariant theories regularized on a lattice. They showed that fermion systems in lattice gauge theories are similar to electron systems in crystals, so as a result there should exist a mechanism in crystals that is similar to the ABJ anomaly. When two energy bands have point-like degeneracies, the Weyl particles can move from one degeneracy to the other in the presence of parallel electric and magnetic fields leading to a large longitudinal positive magnetoconductance~\cite{imran2018berry}, which is a manifestation of the anomaly in crystals. The exact sign of LMC is more nuanced and we discuss this in detail in a later section. 

We present a brief overview of CA with help of a simple Landau-levels picture~\cite{hosur2013recent}.
In Fig.~\ref{fig:WSM_slab_nd_Landau_picture} (a), we sketch a WSM of finite size with volume $V=l_x l_y l_z$, subject to external electric ($\mathbf{E}$) and magnetic field ($\mathbf{B}$). We quantize the levels in presence of an external magnetic field. The  Corresponding energy dispersion is plotted in Fig.~\ref{fig:WSM_slab_nd_Landau_picture} (b), where the energy levels disperse along the direction of magnetic field $\mathbf{B}$. Except $n=0$ level, all the Landau levels are degenerate with degeneracy $g={2\pi e B l_x l_y}/{\hbar}$. The zeroth Landau level is chiral in nature and its direction of dispersion depends upon the chirality $\chi$ of Weyl node. 
The general form of the dispersion may be explicitly written as:
\begin{align}
\epsilon(\mathbf{k})= \nonumber\\
&
\begin{cases}
    v_\mathrm{F} ~\mathrm{sign}(n) \sqrt{2\hbar|n| e B + (\hbar k_z)^2},& n = \pm1, \pm2...\\
    -\chi \hbar v_\mathrm{F} k_z,     & n = 0.
\end{cases}   
\end{align}
We will assume that $\mu \ll v_\mathrm{F} \sqrt{\hbar e B}$. For this case one focuses only on the zeroth Landau level physics; this is also dubbed as the \textit{quantum limit}. 
As depicted in Fig.~\ref{fig:WSM_slab_nd_Landau_picture} (a), for $\mathbf{E} = E \hat{z}$ and $\gamma=\pi/2$, i.e., $\mathbf{B} = B \hat{z}$, all the states are forced to mobilize along $\hat{z}-$direction: ${d \mathbf{k}}/{d t} = ({-e}/{\hbar} )\mathbf{E}$. Therefore, electrons in the zeroth Landau level at valley $\chi= 1$ move towards the left and at valley $\chi= -1$ towards the right. This mobilization of states appears as the disappearance of electrons from the band having a positive slope and reappearance to the band having a negative slope through a hidden channel marked by a black dotted curve in Fig.~\ref{fig:WSM_slab_nd_Landau_picture}(b). 
This process leads to non-conservation of charge at a particular chiral Landau level. In conclusion, chiral Landau levels have 1d chiral anomaly which results in an imbalance of the population of charges at two valleys having opposite chiralities. The rate of change of charge particle at valley having chirality $\chi$ is given by:
\begin{align}
    \frac{\partial \mathcal{Q}^{\chi}_{\hat{z}}}{\partial t} = e \chi l_z \frac{|\dot{\mathbf{k}}|}{2\pi} = -2\pi e^2 \chi l_z \frac{|\mathbf{E}|}{\hbar}.
\end{align} 
Including degeneracy of the Landau levels ($g={2\pi e B l_x l_y}/{\hbar}$), this is generalised as follows:
\begin{align}
    \frac{\partial \mathcal{Q}^{\chi}_{3D}}{\partial t} = g \frac{\partial \mathcal{Q}^{\chi}_{\hat{z}}}{\partial t} = -\frac{e^3}{4\pi^2 \hbar^2} l_x l_y l_z \mathbf{E}\cdot\mathbf{B}.
\end{align}
We will see in the next section how the $\mathbf{E}\cdot\mathbf{B}$ term appears in the semiclassical equations of motion and allows us to study CA and CA-assisted transport through the Bloch-Boltzmann formalism. 

\section{Electron transport in Weyl fermions}
\label{sec2}
\subsection{Maxwell-Boltzmann transport
theory of Weyl fermions}
\label{sec:maxwellboltz}
\begin{figure}
    \centering
    \includegraphics[width=.70\columnwidth]
    {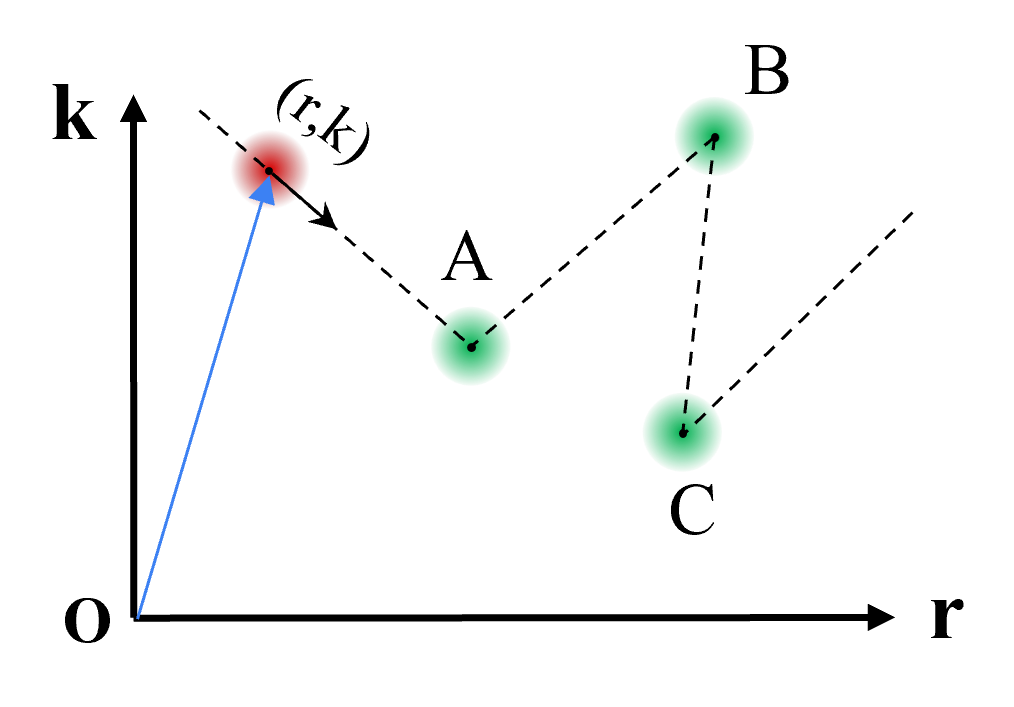}
    \caption{Diagrammatic depiction of the classical route of a Bloch electron wavepacket in the phase space, which is governed by the classical equations of motion i.e., Eq.~\ref{Couplled_equation}. The position is indicated by the blue arrow and can be expressed as the coordinate $(\mathbf{r},\mathbf{k})$. The direction of motion is shown by the black arrow on the dotted route. Impurity sites that are naturally present in the system are denoted by A, B, and C. This semiclassical method enables us to follow an electron between two consecutive collisions.}
\label{fig:Classical_path}
\end{figure}

Electrons in solids are influenced by the periodic lattice potential of ions situated at well-defined basis points. The electronic conductivity of solids therefore must account for this periodic potential. A solution to this problem was given by Bloch~\cite{ashcroft1976nd} and thus noninteracting electrons in periodic potential are dubbed as {Bloch electrons}. Bloch electrons conserve the crystal momentum and have plane-wave solutions modulated by a periodic function: $\psi^n_{\mathbf{k}}(\mathbf{r})=e^{i\mathbf{k}\cdot\mathbf{r}} u^n_\mathbf{k}(\mathbf{r})$, where $n$ is the band index. Extending Sommerfield's theory to nonequilibrium cases in the presence of external perturbations, one then explores the conduction of solids. Here, the dynamics of Bloch electron wavepackets are considered classical. These classical equations describe the behavior of the {wave packet} of electron levels as shown in Fig. \ref{fig:Classical_path}, which is forced to obey the uncertainty principle. 
The equations of motion to track the evolution of the position ($\mathbf{r}$) and wave vector ($\mathbf{k}$) of an electron in an external electromagnetic field ($\mathbf{E}$ and $\mathbf{B}$) are:
\begin{align}
    \mathbf{\dot{r}} \equiv \mathbf{v}=  \frac{1}{\hbar} \frac{\partial \epsilon(\mathbf{k})}{\partial\mathbf{k}},\nonumber\\
    \hbar\mathbf{\dot{k}} = e (\mathbf{E} + \mathbf{\dot{r}} \times \mathbf{B}).
\end{align}
As we noted earlier, Weyl fermions in solids, due to the nontrivial topology of the bands, possess a Berry curvature which modifies the above equation to the one presented in the following equation~\cite{son2012berry}:
\begin{align}
\dot{\mathbf{r}}^\chi &= \mathcal{D}^\chi \left( \frac{e}{\hbar}(\mathbf{E}\times \boldsymbol{\Omega}^\chi) + \frac{e}{\hbar}(\mathbf{v}^\chi\cdot \boldsymbol{\Omega}^\chi) \mathbf{B} + \mathbf{v}_\mathbf{k}^\chi\right) \nonumber\\
\dot{\mathbf{p}}^\chi &= -e \mathcal{D}^\chi \left( \mathbf{E} + \mathbf{v}_\mathbf{k}^\chi \times \mathbf{B} + \frac{e}{\hbar} (\mathbf{E}\cdot\mathbf{B}) \boldsymbol{\Omega}^\chi \right),
\label{Couplled_equation}
\end{align}
where $\boldsymbol{\Omega}^\chi = -\chi \mathbf{k} /2k^3$ is the Berry curvature of Weyl fermions, and $\mathcal{D}^\chi = (1+e\mathbf{B}\cdot\boldsymbol{\Omega}^\chi/\hbar)^{-1}$.  The self-rotation of the Bloch wave packet, which has a finite spread in the phase space, also gives rise to an orbital magnetic moment (OMM) $\mathbf{m}^\chi_\mathbf{k}$~\cite{xiao2010berry}. In the presence of a magnetic field, the OMM shifts the energy dispersion as $\epsilon^{\chi}_{\mathbf{k}}\rightarrow \epsilon^{\chi}_{\mathbf{k}} - \mathbf{m}^\chi_\mathbf{k}\cdot \mathbf{B}$. Note that we have added the chirality index $\chi$ to distinguish Weyl fermions of different flavors. 

Returning to the quasiclassical formalism, it shows that a Bloch electron wavepacket has nonvanishing band velocity ($\mathbf{v}_n$) proportional to $({\partial \epsilon(\mathbf{k})}/{\partial \mathbf{k}} )$. So a perfect solid has infinite conductivity. The inclusion of the wave nature of the electron justifies this as constructive interference of scattered waves from an array of periodic potentials allowing it to propagate through solids without attenuation~\cite{ashcroft1976nd}. Since no crystal structure is perfect, imperfection leads to the degradation of current giving rise to finite conductivity. 
\begin{figure*}
    \centering
    \includegraphics[width=1.9\columnwidth]
    {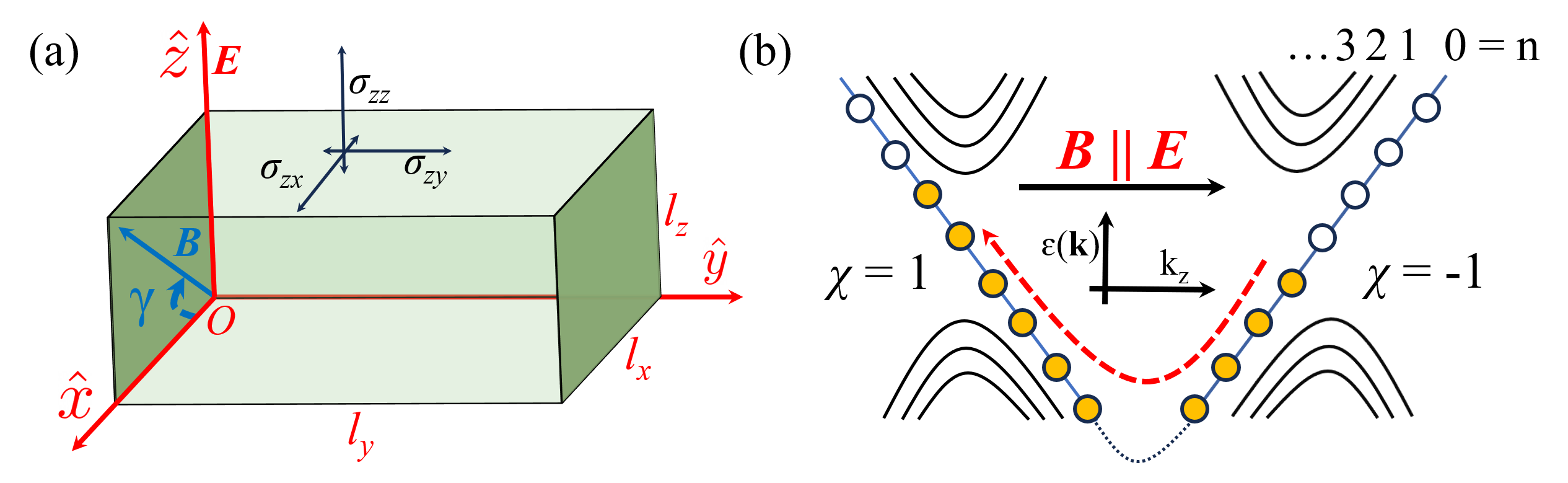}
    \caption{(a) Schematic illustration of the experimental setup to measure different conductivity responses of 3D WSMs under external electric and magnetic fields. $V=l_xl_y l_z$ is sample volume, $\gamma$ is to tune the direction of the magnetic field which is very helpful to study the chiral magnetic effect as chiral current is proportional to $\mathbf{E}\cdot\mathbf{B}$. (b) Landau level picture of the dispersion of WSMs having pair of Weyl nodes with chirality $\chi=\pm1$. Occupancy of the chiral Landau level ($n=0$) has been marked by the yellow circles and it is the only one to participate in the chiral pumping process under parallel electric and magnetic fields.}
\label{fig:WSM_slab_nd_Landau_picture}
\end{figure*}
Using the semiclassical Boltzmann formalism, one can investigate charge transport in the presence of perturbative electric and magnetic fields and evaluate the conductivity. 
For spatially uniform fields, the formalism assumes the presence of a distribution function $f_\mathbf{k}$ of electrons which reduces to the Fermi-Dirac distribution in equilibrium. The nonequilibrium distribution function satisfies the following equation:~\cite{bruus2004many} 
\begin{align}
\dfrac{\partial f_{\mathbf{k}}}{\partial t}+ {\Dot{\mathbf{r}}_{\mathbf{k}}}\cdot \mathbf{\nabla_r}{f_{\mathbf{k}}}+\Dot{\mathbf{k}}\cdot \mathbf{\nabla_k}{f_{\mathbf{k}}}=I_{\mathrm{coll}}[f_{\mathbf{k}}].
\label{MB_equation}
\end{align}
We assume $f_\mathbf{k} = f_{0\mathbf{k}} + g_{\mathbf{k}}$, where $f_{0\mathbf{k}}$ is the standard Fermi-Dirac distribution and $g_{\mathbf{k}}$ is the deviation due to external fields. Perturbation theory allows us to go up to the order of choice to discuss the higher orders of different electromagnetic responses of the materials. When restricted to the first order in $\mathbf{E}$, then $g_\mathbf{k}\propto \mathbf{E}$.
The right-hand side in Eq.~\ref{MB_equation} (the collision integral term) accounts for the relaxation processes in the system. The simplest approach adapted in the literature is the `relaxation time approximation' with momentum-independent constant scattering time that assumes
\begin{align}
    I_{\mathrm{coll}}[f_{\mathbf{k}}] = -\frac{g_\mathbf{k}}{\tau},
\end{align}
where $\tau$ is the scattering time. Once the distribution function is obtained, conductivity can be straightforwardly obtained via the following relation:
\begin{align}
    \mathbf{j} = -e\sum_\mathbf{k} \dot{\mathbf{r}} g_\mathbf{k}.
\end{align}

In the case of Weyl fermions, we need to consider the scattering of an electron from an occupied state with momentum $\mathbf{k}$ and chirality $\chi$ to an unoccupied state with momentum $\mathbf{k}'$ and chirality $\chi'$. This leads to two processes: (i) if $\chi=\chi'$, this is termed as intranode (or intravalley) scattering, referring to the preservation of the chirality index of the particle, (ii) when $\chi\neq\chi'$, this is termed as internode (or intervalley) scattering referring to breaking of the chirality index. Therefore one typically needs to consider two scattering times: (i) $\tau_{\mathrm{inter}}$, which is the internode scattering time, and (ii) $\tau_\mathrm{intra}$, which is the intranode scattering time. Furthermore, the distribution functions at both nodes must have a chirality index as well. The collision integral can therefore be expressed as~\cite{son2013chiral,kim2014boltzmann} 
\begin{align}
    I_{\mathrm{coll}}[f^\chi_{\mathbf{k}}] = -\frac{f^\chi_{\mathbf{k}}-f^{\chi'\neq\chi}_{\mathbf{k}}}{\tau_{\mathrm{inter}}}-\frac{f^\chi_{\mathbf{k}}-f^{\chi'=\chi}_{\mathbf{k}}}{\tau_{\mathrm{intra}}}.
\end{align}
The internode scattering mechanism successfully captures the essence of chiral anomaly-induced transport phenomena as first pointed out by Son and Spivak~\cite{son2013chiral}. It predicts \textit{negative} longitudinal magnetoresistance, which was initially considered a definitive signature of the manifestation of chiral anomaly in solids. Recent works~\cite{knoll2020negative,sharma2020sign,ahmad2021longitudinal,sharma2023decoupling,ahmad2023longitudinal} have evolved this understanding, which we will discuss later.

The linear response formalism for computing conductivity~\cite{mahan20089}, presupposes a timescale denoted by $\tau_{\phi}$, which signifies the interactions between the system and the external electric field. This timescale reflects the inelastic exchange of energy at a rate of $\tau_{\phi}$.
The rate is ideally assumed to be zero, meaning that $\tau_{\phi}$ is presumed to be the longest among all relevant timescales.
For weakly disordered Weyl semimetals, particularly when Landau quantization of energy levels becomes significant under intense magnetic fields, chiral anomaly becomes apparent through a positive contribution to the longitudinal magnetoconductance, expressed as $\mathbf{j}\propto B(\mathbf{E} \cdot \mathbf{B})$. The current flow is constrained by the internode scattering time ($\tau_\mathrm{inter}$), representing the timescale at which electrons scatter across the nodes and alter their chirality. Consequently, while the chiral charge isn't conserved, the global charge is conserved. For the linear response formalism to be valid, $\tau_\mathrm{inter}$ must significantly exceed $\tau_\phi$. However, when intranode scattering is the dominant scattering mechanism, the conservation of chiral charge along with global charge becomes important, and the calculations need to be re-examined.

When magnetic fields are weak and quantization of levels is unimportant, Son and Spivak~\cite{son2013chiral} predicted internode scattering induced positive longitudinal magnetoconductivity (LMC) in Weyl semimetals (WSMs) via the semiclassical Boltzmann approach. Later on, several studies proposed that positive LMC can arise solely from intranode scattering, as evidenced by a coupling term $\mathbf{E}\cdot\mathbf{B}$ (Eq.~\ref{Couplled_equation}) incorporated into the semiclassical equations of motion~\cite{kim2014boltzmann,lundgren2014thermoelectric,cortijo2016linear,sharma2016nernst,zyuzin2017magnetotransport,das2019berry,kundu2020magnetotransport,knoll2020negative}. This suggests that positive LMC can occur in WSMs even in the limit where $\tau_\mathrm{inter}/\tau_\mathrm{intra}\rightarrow\infty$. Notably, none of the studies differentiate between two parameter regimes: $\tau_\mathrm{intra}\ll\tau_\phi\ll\tau_\mathrm{inter}$ and $\tau_\mathrm{intra}\ll\tau_\mathrm{inter}\ll\tau_\phi$. This distinction carries significant consequences, as the chiral charge is conserved in the former case, while the latter indicates global charge conservation, occurring on a timescale larger than the intravalley timescale $\tau_\mathrm{intra}$ but smaller than $\tau_\phi$. The assertion that intranode scattering alone can yield positive LMC  assumes that $\tau_\mathrm{intra}\ll\tau_\phi\ll\tau_\mathrm{inter}$, which is incorrect. This issue was recently resolved in Ref.~\cite{sharma2023decoupling} where Sharma \textit{et al.} calculated transport properties including LMC correctly for various possible values of the parameters $\tau_\mathrm{intra}$, 
 $\tau_\mathrm{inter}$, and $\tau_\phi$. 

Furthermore, in the same work~\cite{sharma2023decoupling}, Sharma \textit{et al.}, showed that a constant relaxation time approximation in Weyl fermions is inherently inconsistent with charge conservation. This is understood by the following illustrative example. Consider one Weyl node with chirality $\chi$ and the following low-energy Hamiltonian: $H_\mathbf{k}=\chi \hbar v_F \mathbf{k}\cdot\boldsymbol{\sigma}$. The steady-state Boltzmann equation takes the following form in the relaxation-time approximation~\cite{sharma2023decoupling}:
\begin{align}
    e \mathcal{D}^\chi_\mathbf{k} \left(-\frac{\partial f_{0\mathbf{k}}}{\partial \epsilon_\mathbf{k}}\right) \left(\mathbf{v}^\chi_\mathbf{k}+ \frac{e}{\hbar}\mathbf{B} (\boldsymbol{\Omega}^\chi_\mathbf{k}\cdot\mathbf{v}^\chi_\mathbf{k})\right)\cdot\mathbf{E} = -\frac{g^\chi_\mathbf{k}}{\tau_\mathbf{k}}
    \label{Eq:Boltz1}
\end{align}
Charge conservation suggests that $
\sum\limits_\mathbf{k}g^\chi_\mathbf{k} = 0$. 
When both $\mathbf{E}$ and $\mathbf{B}$ are parallel to the $z-$axis, the charge conservation equation reduces to:
\begin{align}
    \int {\tau^\chi(\theta) \left({v}^\chi_{z}+ \frac{e}{\hbar}{B} (\boldsymbol{\Omega}^\chi_\mathbf{k}\cdot\mathbf{v}^\chi_\mathbf{k}) \right) \frac{k^3(\theta) \sin\theta}{|\mathbf{v}^\chi_\mathbf{k}\cdot\mathbf{k}|} d\theta} = 0. 
    \label{Eq:particle_consv_2}
\end{align}
Here, all quantities in the integrand are evaluated on the Fermi surface at zero temperature. A common simplification often utilized is to assume that the scattering time is independent of momentum $\mathbf{k}$, denoted as $\tau^\chi(\theta) = \tau^\chi$~\cite{kim2014boltzmann,lundgren2014thermoelectric,cortijo2016linear,sharma2016nernst,zyuzin2017magnetotransport,das2019berry,kundu2020magnetotransport}. However, it can be readily observed from Eq.~\ref{Eq:particle_consv_2} that when $\tau^\chi(\theta)$ is independent of $\theta$, the left-hand side of the equation does not reduce to zero. Therefore, a momentum-independent scattering time is inherently incompatible with particle number conservation. It is important to go beyond the constant relaxation-time approximation, which is reviewed next. 

\subsection{Beyond constant relaxation-time approximation}
\label{sec:beyond rta}
To correctly study magnetotransport, going beyond the constant relaxation-time approximation is inevitable~\cite{knoll2020negative,sharma2020sign,ahmad2021longitudinal,ahmad2023longitudinal,sharma2023decoupling}. We therefore write the collision integral as: 
\begin{align}
 I_{coll}[f^{\chi}_{\mathbf{k}}]=\sum_{\chi' \mathbf{k}'}{{W}^{\chi \chi'}_{\mathbf{k k'}}}{(f^{\chi'}_{\mathbf{k'}}-f^{\chi}_{\mathbf{k}})},
 \label{Collision_integral}
\end{align}
where the scattering rate ${{W}^{\chi \chi'}_{\mathbf{k k'}}}$ is evaluated in the first Born approximation (Fermi's golden rule) as:
\begin{align}
{\mathbf{W}^{\chi \chi'}_{\mathbf{k k'}}} = \frac{2\pi n}{\mathcal{V}}|\bra{u^{\chi'}(\mathbf{k'})}U^{\chi \chi'}_{\mathbf{k k'}}\ket{u^{\chi}(\mathbf{k})}|^2\delta(\epsilon^{\chi'}(\mathbf{k'})-\epsilon_F).
\label{Fermi_gilden_rule}
\end{align}
In the above expression $n$ is the impurity concentration, $\mathcal{V}$ is the system volume, $\ket{u^{\chi}(\mathbf{k})}$ is the Weyl spinor wave function obtained from diagonalization of the Hamiltonian in Eq.~\ref{Eq_H_weyl}, $U^{\chi \chi'}_{\mathbf{k k'}}$ is the scattering potential, and $\epsilon_F$ is the Fermi energy. We choose $U^{\chi \chi'}_{\mathbf{k k'}}$ to model non-magnetic point-like impurities. So, in general $U^{\chi \chi'}_{\mathbf{k k'}} = U^{\chi \chi'} \sigma_{0}$, where $\sigma_0$ is an identity matrix, and the parameter $U^{\chi \chi'}$ can distinguish the intervalley and intravalley scattering. We do not discuss magnetic impurities here but a recent work of ours discusses that as well~\cite{varma2024magnetotransport}.

Using Eq.~\ref{Couplled_equation},~\ref{Collision_integral} and keeping terms up to linear order in the applied fields, the Boltzmann transport equation can be written as:
\begin{align}
&\left[\left(\frac{\partial f_0^\chi}{\partial \epsilon^\chi_\mathbf{k}}\right) \mathbf{E}\cdot \left(\mathbf{v}^\chi_\mathbf{k} + \frac{e\mathbf{B}}{\hbar} (\boldsymbol{\Omega}^\chi\cdot \mathbf{v}^\chi_\mathbf{k}) \right)\right]\nonumber\\
 &= -\frac{1}{e \mathcal{D}^\chi}\sum\limits_{\chi'}\sum\limits_{\mathbf{k}'} W^{\chi\chi'}_{\mathbf{k}\mathbf{k}'} (g^\chi_{\mathbf{k}'} - g^\chi_\mathbf{k}).
 \label{Eq_boltz2}
\end{align}
This is a vector equation that can be simplified by fixing the electric field along a particular direction. We fix the electric field along increasing $x$-direction and the magnetic field can be rotated in $xz$-plane. Therefore, $\mathbf{E} = E(0,0,1)$ and  $\mathbf{B} = B (\cos{\gamma},0,\sin{\gamma})$, i.e., for $\gamma=\pi/2$ both the fields are parallel. By tuning the $\gamma$ we can control the component of the magnetic field along the electric field, i.e. $\mathbf{E}\cdot\mathbf{B}$ term in the Eq.~\ref{Couplled_equation} which is responsible for the chiral anomaly. This geometry is useful for studying the different electromagnetic responses, especially the planar Hall effect and longitudinal magnetoconductance (we will return to this later). In this geometry, only the $z$-component of $\mathbf{\Lambda}^{\chi}_{\mathbf{k}}$  is relevant, so Eq.~\ref{Eq_boltz2} reduces to
\begin{align}
\mathcal{D}^{\chi}\left[{v^{\chi,z}_{\mathbf{k}}}+\frac{eB\sin{\gamma}}{\hbar}(\mathbf{v^{\chi}_k}\cdot\mathbf{\Omega}^{\chi}_k)\right] \sum_{\chi' \mathbf{k}'}{{W}^{\chi \chi'}_{\mathbf{k k'}}}{(\Lambda^{\chi'}_{\mathbf{k'}}-\Lambda^{\chi}_{\mathbf{k}})}.
\label{boltzman_in_terms_lambda}  
\end{align}
We define a valley scattering rate:
\begin{align}
\frac{1}{\tau^{\chi}_{\mathbf{k}}(\theta,\phi)}=\sum_{\chi'}\mathcal{V}\int\frac{d^3\mathbf{k'}}{(2\pi)^3}(\mathcal{D}^{\chi'}_{\mathbf{k}'})^{-1}{W}^{\chi \chi'}_{\mathbf{k k'}}.
\label{Tau_invers}
\end{align}
On the right-hand side of the above equation, there is a sum over the chirality index $\chi$ which can run over multiple flavors. For example, time reversal symmetry broken WSM has a minimum of two Weyl cones, so the summation is over two nodes. But, inversion symmetry broken WSM has a minimum of four Weyl cones so it runs for all four nodes.  The overlap of the Bloch wave-functions $\bra{u^{\chi'}(\mathbf{k'})}U^{\chi \chi'}_{\mathbf{k k'}}\ket{u^{\chi}(\mathbf{k})}|^2$ in Eq.~\ref{Fermi_gilden_rule} is given by the following expression,
$\mathcal{G}^{\chi\chi'}(\theta,\phi) = [1+\chi\chi'(\cos{\theta}\cos{\theta'} + \sin{\theta}\sin{\theta'}\cos{\phi}\cos{\phi'} + \sin{\theta}\sin{\theta'}\sin{\phi}\sin{\phi'}]$. The overlap of the wave function includes the transition probabilities between fermions of the same chiralities as well as different chiralities. Taking Berry phase into account and the corresponding change in the density of states, $\sum_{k}\longrightarrow \mathcal{V}\int\frac{d^3\mathbf{k}}{(2\pi)^3}\mathcal{D}^\chi(k)$, Eq.~\ref{boltzman_in_terms_lambda} becomes:
\begin{align}
h^{\chi}_{\mu}(\theta,\phi) + &\frac{\Lambda^{\chi}_{\mu}(\theta,\phi)}{\tau^{\chi}_{\mu}(\theta,\phi)}=\nonumber\\
&\sum_{\chi'}\mathcal{V}\int\frac{d^3\mathbf{k}'}{(2\pi)^3} \mathcal{D}^{\chi'}(k'){W}^{\chi \chi'}_{\mathbf{k k'}}\Lambda^{\chi'}_{\mu}(\theta',\phi')
\label{MB_in_term_Wkk'}
\end{align}
Here, the explicit form of the $h^{\chi}_{\mu}$ is $h^{\chi}_{\mu}(\theta,\phi)=\mathcal{D}^{\chi}[v^{\chi}_{z,\mathbf{k}}+eB\sin{\gamma}(\mathbf{\Omega}^{\chi}_{k}\cdot \mathbf{v}^{\chi}_{\mathbf{k}})]$ and is independent of the nature of the impurity sites. The momentum integral in Eq.~\ref{MB_in_term_Wkk'} has to be evaluated at the Fermi surface, and for that one has to change the momentum integration into the energy integration. 
In the zero temperature limit, for a constant Fermi energy surface, the Eq.~\ref{Tau_invers} and RHS of Eq.~\ref{MB_in_term_Wkk'} is reduced to the integration over $\theta'$ and $\phi'$:
\begin{align}
\frac{1}{\tau^{\chi}_{\mu,i}(\theta,\phi)} =  \mathcal{V}\sum_{\chi'} \Pi^{\chi\chi'}\iint\frac{(k')^3\sin{\theta'}}{|\mathbf{v}^{\chi'}_{k'}\cdot{\mathbf{k'}^{\chi'}}|}d\theta'd\phi' \mathcal{G}^{\chi\chi'}(\mathcal{D}^{\chi'}_{\mathbf{k'}})^{-1}
\label{Tau_inv_int_thet_phi}
\end{align}
The parameter $\tau$ in this expression is independent of radial distance but is dependent on the angles $\theta$,$\phi$, and the nature of the impurity sites.  By using this expression for $\tau$, and Eq.~\ref{Fermi_gilden_rule} the right-hand side of the Eq.~\ref{MB_in_term_Wkk'} takes the form:
\begin{multline}
[d^{\chi}+a^{\chi}\cos{\phi}+b^{\chi}\sin{\theta}\cos{\phi}+c^{\chi} \sin{\theta}\sin{\phi}]\\
=\sum_{\chi'}\mathcal{V}\Pi^{\chi\chi'}\iint f^{\chi'}(\theta',\phi')d\theta'd\phi'\\\times[d^{\chi'}-h^{\chi'}_{k'}+a^{\chi'}\cos{\theta'}+b^{\chi'}\sin{\theta'}\cos{\phi'}+c^{\chi'} \sin{\theta'}\sin{\phi'}],\\
\label{Boltzman_final}
\end{multline}
where $\Pi^{\chi \chi'} = N|U^{\chi\chi'}|^2 / 4\pi^2 \hbar^2$, $f^{\chi} (\theta,\phi)=\frac{(k)^3}{|\mathbf{v}^\chi_{\mathbf{k}}\cdot \mathbf{k}^{\chi}|} \sin\theta (\mathcal{D}^\eta_{\mathbf{k}})^{-1} \tau^\chi_\mu(\theta,\phi)$ and the exact form of the ansatz is $\Lambda^{\chi}_{\mathbf{k}}=[d^{\chi}-h^{\chi}_{k'} + a^{\chi}\cos{\phi} +b^{\chi}\sin{\theta}\cos{\phi}+c^{\chi}\sin{\theta}\sin{\phi}]\tau^{\chi}_{\mu}(\theta,\phi)$. 
When the aforementioned equation is explicitly put out, it appears as seven simultaneous equations that must be solved for eight variables.  
The particle number conservation provides the final restriction:
\begin{align}
\sum\limits_{\chi}\sum\limits_{\mathbf{k}} g^\chi_\mathbf{k} = 0.
\label{Eq_sumgk}
\end{align} 
For the eight unknowns ($d^{\pm 1}, a^{\pm 1}, b^{\pm 1}, c^{\pm 1}$), the equations \ref{Boltzman_final} and \ref{Eq_sumgk} are simultaneously solved with Eq.~\ref{Tau_inv_int_thet_phi}. Note that if we have more than two flavors of Weyl fermions, the number of equations and unknowns will increase. For example, in the case of inversion asymmetric WSMs, we need to numerically solve sixteen equations for sixteen coefficients~\cite{ahmad2021longitudinal,sharma2023decoupling}. 

\section{Anomalous transport responses}
\label{sec:sec3}
\subsection{Longitudinal magnetoconductace}
\label{sec:lmc}
Intial calculations suggested that CA always resulted in a positive LMC~\cite{fukushima2008chiral,dos2016search,wu2021valley,lucas2016hydrodynamic, son2013chiral,spivak2016magnetotransport} or equivalently, negative magnetoresistance. Son~\& Spivak~\cite{son2013chiral} showed that a Weyl metal, that has Weyl points immersed inside the Fermi surface can have a chiral anomaly induced large negative magnetoresistance ~\cite{son2013chiral}. The explanation offered was semiclassical where the mean free path is short compared to the magnetic length and Landau quantization of levels is irrelevant. As a result, the electron transport properties were studied using the Boltzmann equation, predicting the electric conductivity to depend quadratically on the applied magnetic field. Burkov~\cite{burkov2014chiral} arrived at the same conclusion using a microscopic theory of diffusive transport. A later work by Spivak~\& Andreev ~\cite{spivak2016magnetotransport} also explained the occurrence of large negative magnetoconductance in Weyl metals due to CA using the semiclassical Boltzmann transport theory. According to them, negative magnetoresistance in a conductor in weak fields can easily be explained with Boltzmann transport theory by invoking Berry curvature without which the magnetoresistance remains positive. A lot of experimental observations of possible WSM candidates have shown large magnetoresistance consistent with the above studies~\cite{he2014quantum,liang2015ultrahigh,li2016chiral,xiong2015evidence,hirschberger2016chiral}. While correct in spirit, these theoretical works assume a momentum-independent relaxation time, which as we show earlier, is an incorrect assumption for Weyl fermions. 

It has come to light that positive longitudinal magnetoconductance doesn't exclusively confirm the presence of chiral anomaly in Weyl semimetals. Notably, positive LMC can manifest even without Weyl nodes, as observed in ultraclean PdCoO$_2$~\cite{kim2009fermi,noh2009anisotropic,kikugawa2016interplanar}. Furthermore, experimental studies indicate that the jetting effect can yield false positive LMC due to extrinsic factors, irrespective of the presence of any anomaly~\cite{dos2016search}. However, a recent work~\cite{liang2018experimental} demonstrates that meticulous measurement of voltage drops along the sample's mid ridge and edges can mitigate this effect. Theoretically, it is now also established that Weyl nodes don't invariably lead to positive LMC. In strong magnetic fields, where one needs to consider effects due to Landau quantization, LMC can be both positive or negative for short-range impurities, while it is typically positive for long-range charged impurities~\cite{goswami2015axial,lu2015high,chen2016positive,zhang2016linear,shao2019magneto,li2016weyl,ji2018effect}. In the regime of weak magnetic fields, it was conventionally understood that LMC is positive~\cite{spivak2016magnetotransport,das2019linear,imran2018berry,dantas2018magnetotransport,johansson2019chiral,grushin2016inhomogeneous,cortijo2016linear,sharma2017chiral}. However, a recent works~\cite{knoll2020negative, sharma2020sign, sharma2023decoupling, ahmad2023longitudinal,ahmad2021longitudinal} reveal that LMC can be negative with sufficiently strong intervalley scattering, especially when the effects of orbital magnetic moment (OMM) and are considered beyond the constant relaxation-time approximation. 

Knoll \textit{et al.}~\cite{knoll2020negative} studied the CA-induced LMC in WSMs using the semiclassical theory, including the effects of the anomalous orbital magnetic moment and going beyond the constant relaxation-time approximation. They explain that the anomalous shift in the energy due to the OMM, which changes the geometry of the Fermi surface from spherical to egg-shape, along with the relative strength of the intervalley scattering, is crucial in determining the sign of LMC in the weak-field limit. For small values of relative intervalley scattering strength, LMC is found positive. If the relative intervalley scattering strength is increased further, there is a sign change of LMC from positive to negative at some critical value of $\alpha$. This is also highlighted in Fig.~\ref{fig:two_node_1}. Similar conclusions were reported by Sharma \textit{et al.} in Ref.~\cite{sharma2020sign}. %

\subsubsection{Single node LMC}
Sharma \textit{et al.}~\cite{sharma2023decoupling} revisited this problem and first evaluated LMC for strictly zero internode scattering. We briefly summarize this here. Before proceeding with the calculation, we point out that we adopt the following Weyl Hamiltonian near a nodal point of chirality $\chi$:
\begin{align}
H_\mathbf{k}=\chi\hbar v_F \boldsymbol{\sigma}\cdot\mathbf{k},
\end{align}
where we have inserted the factors of Fermi velocity and the reduced Planck's constant to be dimensionally consistent.

In this scenario, it is enough to compute the outcome for just one isolated Weyl node, and then sum the contributions of both the nodes. Given the inadequacy of the momentum-independent relaxation time, we opt for the collision integral to be:
\begin{align}
    \mathcal{I}_\mathrm{coll}[g_\mathbf{k}^\chi]=\sum\limits_{\mathbf{k}'} \left(\Lambda^\chi_\mathbf{k} - \Lambda^\chi_{\mathbf{k}'}\right) W_{\mathbf{k}\mathbf{k}'}^{\chi\chi}{\left(-\partial g_0/\partial\epsilon_k\right) eE}.
\end{align}
The unknown function $\Lambda^\chi_\mathbf{k}$ is assumed to be 
\begin{align}
    \Lambda^\chi_\mathbf{k} = (f^\chi_\mathbf{k} - h_\mathbf{k}^\chi) \tau_\mathbf{k}^\chi,
\end{align}
where $h^\chi_\mathbf{k} = \mathcal{D}^\chi_\mathbf{k} \left({v}^\chi_z+ {e}{\hbar^{-1}}{B} (\boldsymbol{\Omega}^\chi_\mathbf{k}\cdot\mathbf{v}^\chi_\mathbf{k})\right)$, $(\tau^\chi_\mathbf{k})^{-1} = \sum_{\mathbf{k}'} W_{\mathbf{k}\mathbf{k}'}^{\chi\chi}$, and   $f_\mathbf{k}^\chi$ is the new unknown function. The role of chemical potential and the orbital magnetic moment enter $\left(-\partial g_0/\partial\epsilon_k\right)$ and a $B$-dependent energy shift in the dispersion, respectively. Using the Fermi-golden rule and the simplest scenario of point-like non-magnetic disorder, the scattering rate is determined by
\begin{align}
    W_{\mathbf{k}\mathbf{k}'} &= \frac{2\pi}{\hbar} |U|^2 \delta(\epsilon_\mathbf{k} - \epsilon_F) \times \nonumber\\&(1+\cos\theta\cos\theta' + \sin\theta\sin\theta' \cos(\phi-\phi')).
    \label{Eq_W_1}
\end{align}
Here, $U$ represents the disorder strength in suitable units. It is crucial to highlight that the explicit dependence on momentum direction, namely the polar and azimuthal angles, arises from the chirality of the Weyl fermion wavefunction. This dependence persists even in the absence of a momentum-dependent disorder strength $U$. Consequently, the Boltzmann equation simplifies to
\begin{align}
    f^\chi(\theta) =  \int{F^\chi(\theta')\tau(\theta) (1+\cos\theta\cos\theta') (f^\chi(\theta') - h^\chi(\theta')) d\theta'},
\end{align}
where $F^\chi(\theta) = k^3(\theta) (\mathcal{D}^{\chi}_\theta)^{-1} \sin\theta  |\mathbf{v}_\mathbf{k}\cdot\mathbf{k}|^{-1}$. All the quantities above are evaluated on the Fermi surface. Incorporating particle conservation by employing the ansatz below solves the aforementioned Boltzmann transport equation.
\begin{align}
    \Lambda^\chi(\theta) = \tau(\theta) (a^\chi + b^\chi \cos\theta - h^\chi(\theta)).
\end{align}
 Finally, the current is evaluated as: 
\begin{align}
    \mathbf{j}^\chi = -e\sum\limits_{\mathbf{k}}{\dot{\mathbf{r}}^\chi ( g^\chi_\mathbf{k})},
\end{align}
and thus the longitudinal magnetoconductivity $\sigma_{zz}(B)$ is evaluated. We define $\delta\sigma_{zz}(B) = \sigma_{zz}(B) - \sigma^0_{zz}$, where $\sigma^0_{zz}$ is the zero field conductivity.
\begin{figure}
    \centering
    \includegraphics[width=\columnwidth]{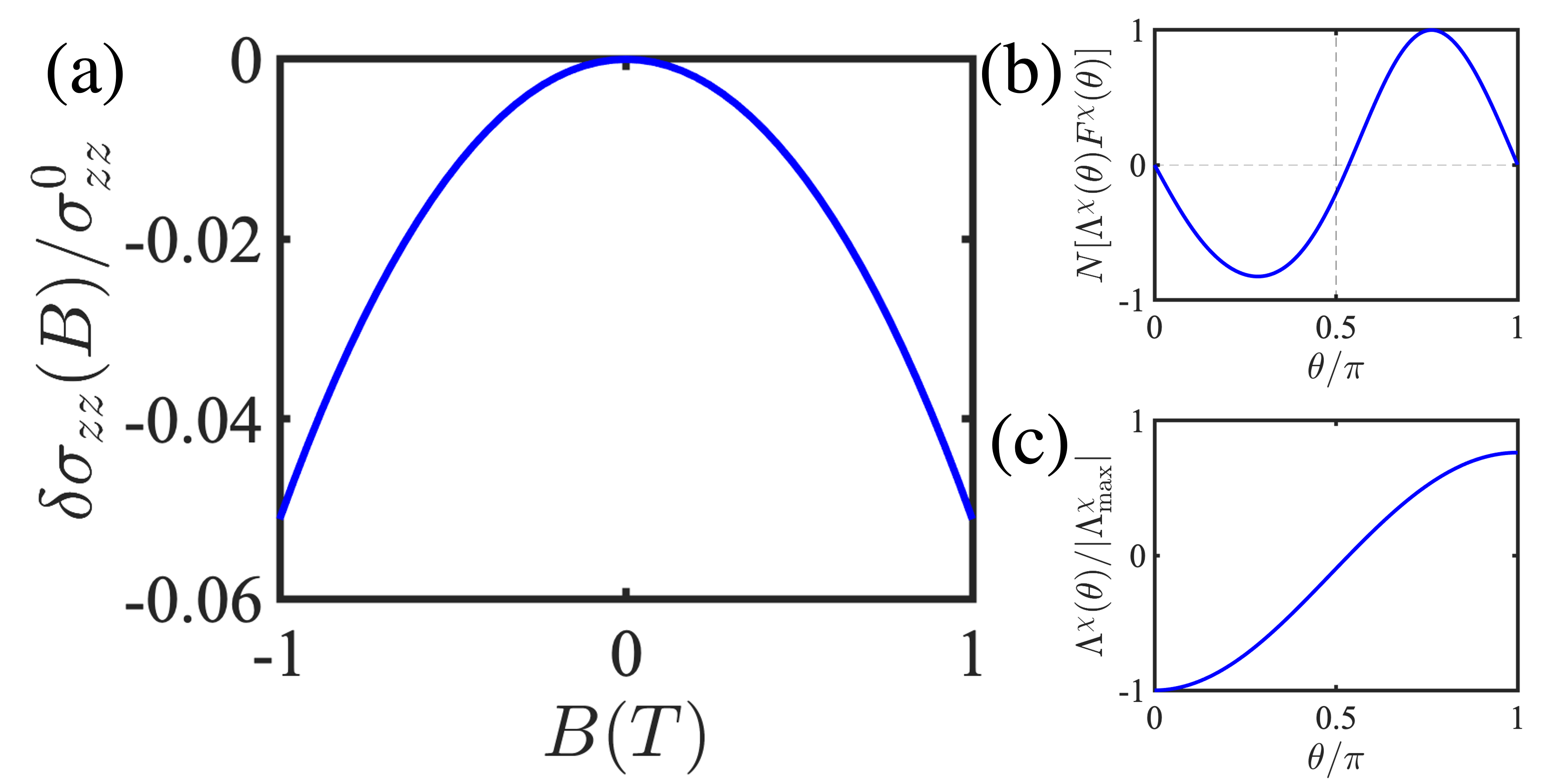}
    \caption{(a) LMC for an isolated Weyl node is always negative due to chiral charge conservation. (b)  $\Lambda^\chi(\theta)$ between $0$ and $\pi$ at $B=1T$. (c) The corresponding normalized deviation in the distribution function (proportional to $g^\chi_\mathbf{k}$), which integrates to zero. Here $\chi=1$. Figure adapted from Ref.~\cite{sharma2023decoupling}.}
    \label{fig:one_node_1}
\end{figure}

Figure \ref{fig:one_node_1} (a) illustrates the computed Longitudinal magnetoconductivity coefficient (LMC), $\delta\sigma_{zz}(B)$, for an isolated Weyl node. Remarkably, it consistently exhibits negative LMC, regardless of the Weyl node's chirality. This finding is in sharp contrast to claims from recent studies that suggest that intranode scattering alone could yield positive LMC in Weyl semimetals \cite{kim2014boltzmann,lundgren2014thermoelectric,cortijo2016linear,sharma2016nernst,zyuzin2017magnetotransport,knoll2020negative,das2019berry,kundu2020magnetotransport}. Figure \ref{fig:one_node_1} (b) illustrates $\Lambda^\chi(\theta)$, while Figure \ref{fig:one_node_1} (c) displays the normalized distribution function. Notably, the distribution function directly reflects $g^\chi_\mathbf{k}$, which integrates to zero, aligning with particle number conservation. Additionally, Ref.~\cite{sharma2023decoupling} also tackles this problem using an ansatz-free numerical solution to the Boltzmann equation yielding consistent results.

\subsubsection{Single node LMC from Landau levels} 
Here we briefly review the magnetoconductivity calculation using the Landau level formalism for an isolated single node that is also performed in Ref~\cite{sharma2023decoupling}.
The magnetic field is fixed along the $z-$axis, and the following gauge $\mathbf{A}=(0,Bx,0)$ is chosen. After Peierls substitution, $\mathbf{k}\rightarrow \boldsymbol{\Pi} = \mathbf{k} + e\mathbf{A}/\hbar$, the following annihilation and creation operators are introduced: 
\begin{align}
    a &= \frac{l_B}{\sqrt{2}} (\Pi_x - i \Pi_y) \nonumber\\
    a^\dagger & = \frac{l_B}{\sqrt{2}} (\Pi_x + i \Pi_y),
\end{align}
where $l_B = \sqrt{\hbar/eB}$ is the magnetic length. The Hamiltonian for a single Weyl node is rewritten in terms of the new operators as:
\begin{align}
    H = \hbar v_F \begin{pmatrix}
k & \frac{\sqrt{2}}{l_B} a\\
\frac{\sqrt{2}}{l_B} a^\dagger & -k
\end{pmatrix},
\end{align}
where $k\equiv k_z$, since $k_x$, $k_y$ are no longer good quantum numbers. 
The energy spectrum at a single Weyl node is $\epsilon_{n}(k) = v_F\sqrt{2\hbar neB + (\hbar k)^2}$, and additionally, there is a zeroth quantum level that is given by $\epsilon_0 = -\chi \hbar v_F k$.
The Landau level wavefunctions are given by:
\begin{align}
|\psi_k\rangle = \frac{1}{\sqrt{\mathcal{N}_k}}
    \begin{pmatrix}
    \frac{\sqrt{2(n+1)}}{l_B} \frac{1}{\epsilon_{n}(k)/\hbar v_F - k} |n\rangle \\
    |n+1\rangle
    \end{pmatrix},
\end{align}
where
\begin{align}
    \mathcal{N}_k = 1+ \frac{2(n+1)}{l_B^2}\frac{1}{\left(\epsilon_{n}(k)/\hbar v_F - k\right)^2}.
\end{align}
\begin{figure}
    \centering
    \includegraphics[width=0.85\columnwidth]{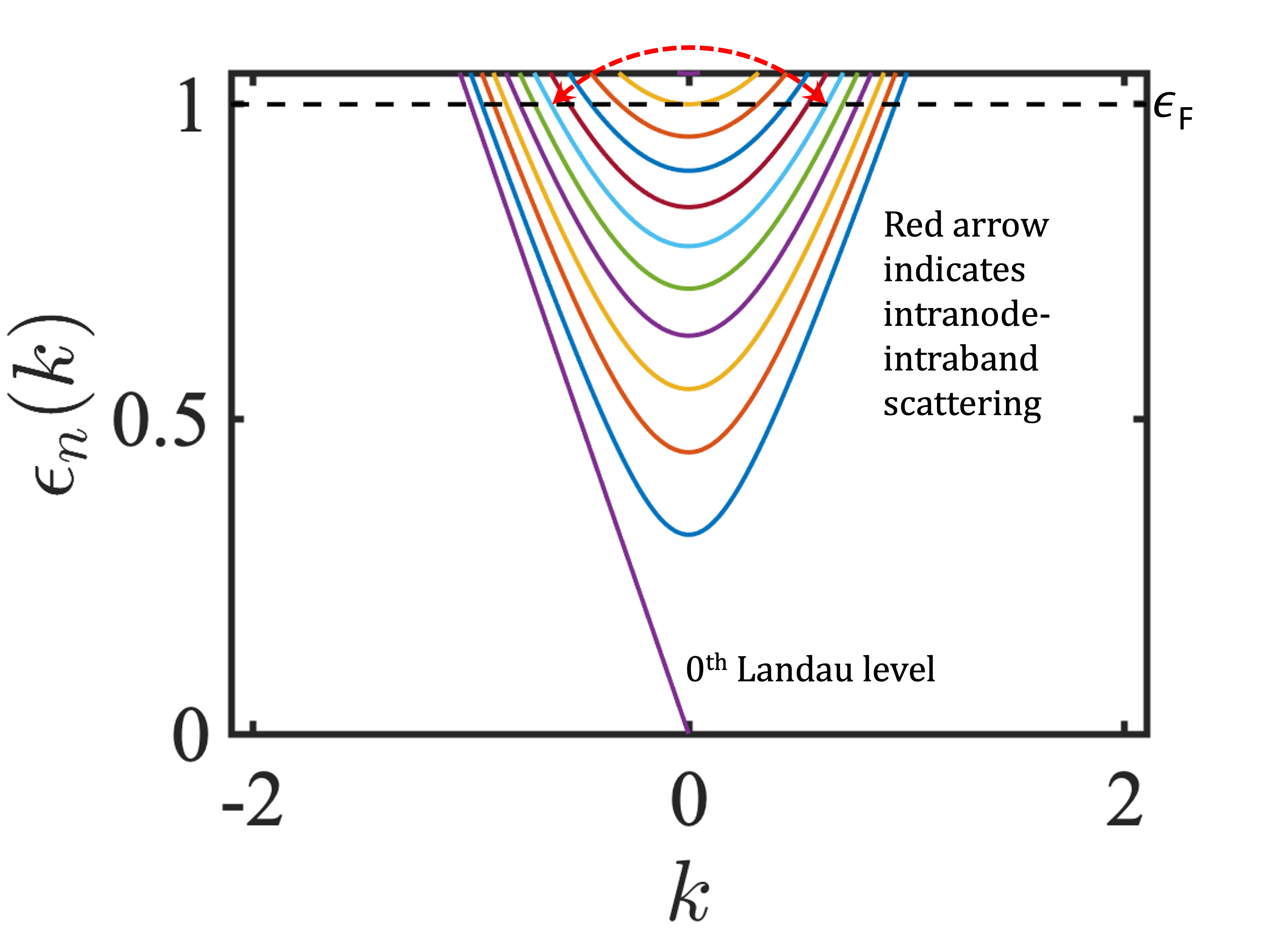}
    \caption{The Landau level spectrum on a single Weyl node. The red arrow indicates the intranode scattering process within the same Landau band (intraband). The zeroth Landau level has no phase space to scatter other than the second node because it is forbidden in the intranode process. Figure adapted from Ref.~\cite{sharma2023decoupling}.}
    \label{fig:llspectrum}
\end{figure}
In Fig.~\ref{fig:llspectrum}, the Landau level spectrum is illustrated at a constant magnetic field strength, varying with $k$. In subsequent calculations, a fixed electron density ($\eta$) is maintained, which is usually the case in experiments. The expression for $\eta$ is defined as:
\begin{align}
\eta = \left(\frac{AeB}{\hbar}\right)\sum\limits_{n=0}^{n_c} \int{\rho^n(\epsilon) f^n(\epsilon) d\epsilon},
\end{align}
where $n$ represents the Landau level index, $n_c$ denotes the number of occupied Landau levels, $\rho^n(\epsilon)$ characterizes the density of states of the $n^\mathrm{th}$ band, $f^n(\epsilon)$ denotes its occupancy, and $AeB/\hbar$ is the degeneracy of each level. The expression for the density of states is given by:
\begin{align}
    \rho^n(\epsilon) = \int{\frac{dk}{2\pi} \delta(\epsilon - \epsilon_n(k))}.
\end{align}
\begin{align}
    \rho^{n\geq 1} (\epsilon) =  \frac{1}{2\pi \hbar v_F^2} \frac{\epsilon}{\sqrt{\epsilon^2/v_F^2 - 2\hbar neB}},
\end{align}
\begin{align}
    \rho^0(\epsilon) = \frac{1}{2\pi \hbar v_F}.
\end{align}
The electronic density $\eta$ then becomes
\begin{align}
    \eta = \left(\frac{AeB}{\hbar}\right)\frac{1}{2\pi \hbar}\left[\sum\limits_{n=1}^{n_c} \sqrt{\frac{\epsilon_F^2}{v_F^2} - 2\hbar n e B}+\frac{\epsilon_F}{v_F}\right].
\end{align}
Since the Fermi energy $\epsilon_F = v_F\sqrt{2 \hbar n_c e B}$, we have 
\begin{align}
        \eta = \left(\frac{AeB}{\hbar}\right)\frac{\sqrt{2 \hbar e B}}{2\pi \hbar}\left[\sum\limits_{n=1}^{n_c} \sqrt{n_c-n}+\frac{\epsilon_F}{v_F}\right]
\end{align}
Evaluating the summation above, we find
\begin{align}
        \eta = \left(\frac{AeB}{\hbar}\right)\frac{\sqrt{2 \hbar e B}}{2\pi \hbar}\left[-H_z(-1/2,n_c)+Z(-1/2)+\frac{\epsilon_F}{v_F}\right].
\end{align}
Here $H_z(s,x)$ is the Hurwitz-Zeta function and $Z(x)$ is the  Riemann-Zeta function. The above equation solves the number for filled Landau levels $n_c$ as a function $B$. 
\begin{figure}
    \centering
    \includegraphics[width=0.85\columnwidth]{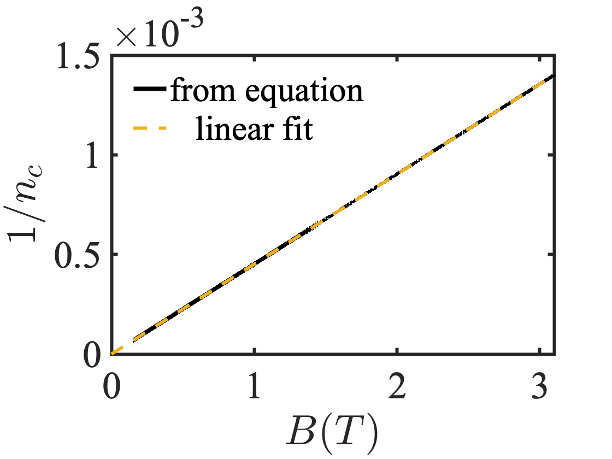}
    \caption{$1/n_c$ as a function of the magnetic field, where $n_c$ is the number of occupied Landau levels. Figure adapted from Ref.~\cite{sharma2023decoupling}.}
    \label{fig:nc_sigma_vs_b}
\end{figure}
Since the analytical expression for $n_c$ is unavailable, it is calculated numerically by inverting the above equation. 
In Fig.~\ref{fig:nc_sigma_vs_b},  $1/n_c$ is plotted as a function of the magnetic field that is observed to be perfectly linear in $B$ for a wide range of magnetic fields. 
The conductivity for each Landau band $n\geq 1$ is given by 
\begin{align}
    \sigma^n = \frac{e^2}{\hbar}\int{\frac{dk}{2\pi} (v_n^z({k}))^2\hspace{1mm} \tau^n_{k}\hspace{1mm} \delta(\epsilon_n({k})-\epsilon_F)},
\end{align}
where 
\begin{align}
    v_n^z({k}) = \frac{1}{\hbar} \frac{\partial \epsilon_n({k})}{\partial k} = \frac{\hbar k v_F^2}{\epsilon_n({k})},
\end{align}
and $\tau^n_{k}$ is the \textit{intranode}, \textit{intra-(Landau)-band} scattering time. For point-like non-magnetic impurities this is evaluated as 
\begin{align}
    \frac{1}{\tau^n_{k}} = \frac{n_\mathrm{imp}V_\mathrm{dis}^2}{\hbar}\int{{dk'}|\langle \psi_k|\psi_{k'} \rangle|^2 \delta(\epsilon_n(k')-\epsilon_F)}.
\end{align}
Here $V_\mathrm{dis}$ is the disorder strength  and $n_\mathrm{imp}$ is the  impurity density.

Before proceeding ahead, it is crucial to highlight the nature of scattering within a single Weyl node and a single Landau band, as depicted in Fig.~\ref{fig:llspectrum}. Specifically, the red arrow's direction illustrates the sole possible scattering channel within a Landau band that conserves energy (elastic scattering). {Notably, the zeroth Landau level lacks scattering phase space within the same node, thereby not contributing to intranode scattering.} However, the opposite-chirality Weyl node provides the sole available scattering phase space for the zeroth Landau level (internode scattering). Consequently, we must discount the zeroth Landau level's contribution when calculating the intranode conductivity. The intranode, intra-(Landau)-band scattering time is determined to be
\begin{align}
    {\tau^n_{k}} = \frac{\hbar^2 v_F}{n_\mathrm{imp}V_\mathrm{dis}^2}\frac{\sqrt{\epsilon_F^2 - 2\hbar n e B v_F^2}}{\epsilon_F}
\end{align}
The total intranode conductivity is given by 
\begin{align}
    \sigma = \sum\limits_{n=1}^{n_c} \sigma^n.
\end{align}
Including the degeneracy, we evaluate $\sigma^n$ to be 
\begin{align}
    \sigma^n = \frac{e^2}{2\pi\hbar}  \frac{AeB}{\hbar} \frac{\hbar v_F^2}{n_\mathrm{imp}V_\mathrm{dis}^2} \frac{{\epsilon_F^2 - 2\hbar n e B v_F^2}}{\epsilon_F^2}.
\end{align}
Substituting $\epsilon_F = v_F\sqrt{2 \hbar n_c e B}$, and evaluating the summation, we find (all the constant prefactors factors are absorbed in $\mathcal{C}$)
\begin{align}
    \sigma = \mathcal{C} B \sum\limits_{n=1}^{n_c} \left(1-\frac{n}{n_c}\right)  =\mathcal{C}B \left(\frac{n_c-1}{2}\right) =\frac{\mathcal{C}}{2}\left({Bn_c}- {B}\right)
\end{align}
Since $Bn_c$ is a constant (from Fig.~\ref{fig:nc_sigma_vs_b}), we conclude that the intranode conductivity decreases with an increase in the magnetic field. 
\begin{figure}
    \centering
    \includegraphics[width=\columnwidth]{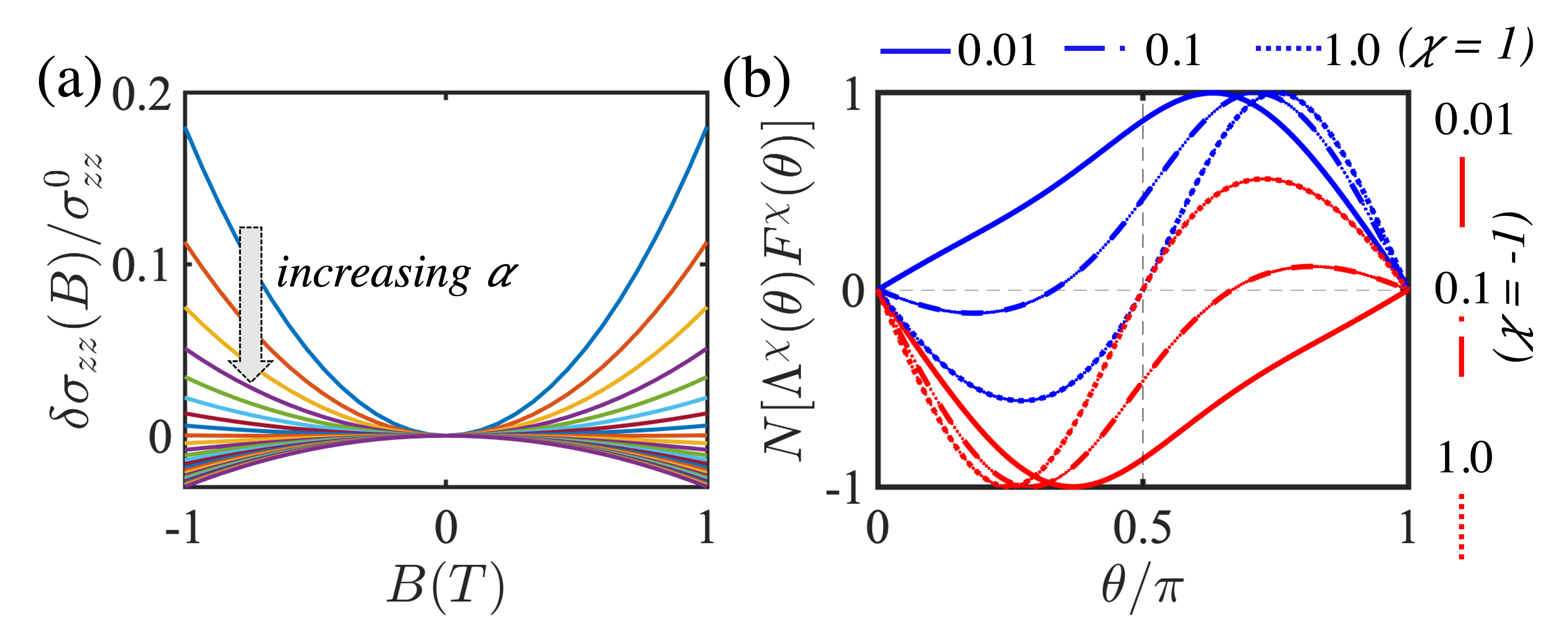}
    \caption{(a) LMC for a pair of Weyl nodes of opposite chiralities becomes negative beyond a critical intervalley scattering strength $\alpha_c$. Here $\alpha$ is increased from 0.1 to 1. (b) The normalized deviation in distribution functions at both the valleys for three different values of $\alpha$. Figure adapted from Ref.~\cite{sharma2023decoupling}.}
    \label{fig:two_node_1}
\end{figure}
\begin{figure}
    \centering
    \includegraphics[width=0.9\columnwidth]{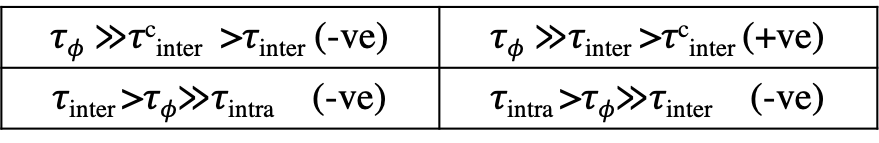}
    \caption{Conditions to observe positive or negative LMC. Figure adapted from Ref.~\cite{sharma2023decoupling}.}
    \label{fig:table}
\end{figure}
\subsubsection{Two-node LMC with internode scattering:}
The previous two subsections have established the fact that the longitudinal magnetoconductivity from a single Weyl node is negative. Now, returning to the case of a pair of Weyl fermions and the Boltzmann formalism, the collision integral takes the following form 
\begin{align}
    \mathcal{I}_\mathrm{coll}[f_\mathbf{k}^\chi]=\sum\limits_{\mathbf{k}'\chi'} \left(\Lambda^{\chi}_\mathbf{k} - \Lambda^{\chi'}_{\mathbf{k}'}\right) W^{\chi\chi'}_{\mathbf{k}\mathbf{k}'} {\left(-\partial g_0/\partial\epsilon_k\right) eE},
\end{align}
where the scattering rate $W^{\chi\chi'}_{\mathbf{k}\mathbf{k}'}$ has both internode and intranode scattering, and is evaluated for non-magnetic pointlike impurities to be:
\begin{align}
    W^{\chi\chi'}_{\mathbf{k}\mathbf{k}'} &= \frac{2\pi}{\hbar} |U^{\chi\chi'}|^2 \delta(\epsilon_\mathbf{k} - \epsilon_F) \times \nonumber\\&(1+\chi\chi'(\cos\theta\cos\theta' + \sin\theta\sin\theta' \cos(\phi-\phi'))).
    \label{Eq_W_2}
\end{align}
Here, the disorder parameter $U^{\chi\chi'}$ allows us to control internode and intranode scattering strengths separately. The ratio $|U^{\chi\neq\chi'}|^2/ |U^{\chi=\chi'}|^2\equiv\alpha$, which is the relative intervalley scattering strength. 
In Figure \ref{fig:two_node_1} (a), we plot LMC against increasing intervalley scattering strength. Remarkably, beyond a critical intervalley scattering strength $\alpha_c$, the LMC is negative. This trend was reported in Refs. \cite{knoll2020negative, sharma2020sign, sharma2023decoupling}, and was attributed to (i) the orbital magnetic moment's opposing effects on the Fermi surfaces of both nodes, and (ii) discarding of the constant relaxation-time approximation.
\begin{figure}
    \centering
    \includegraphics[width=0.9\columnwidth]{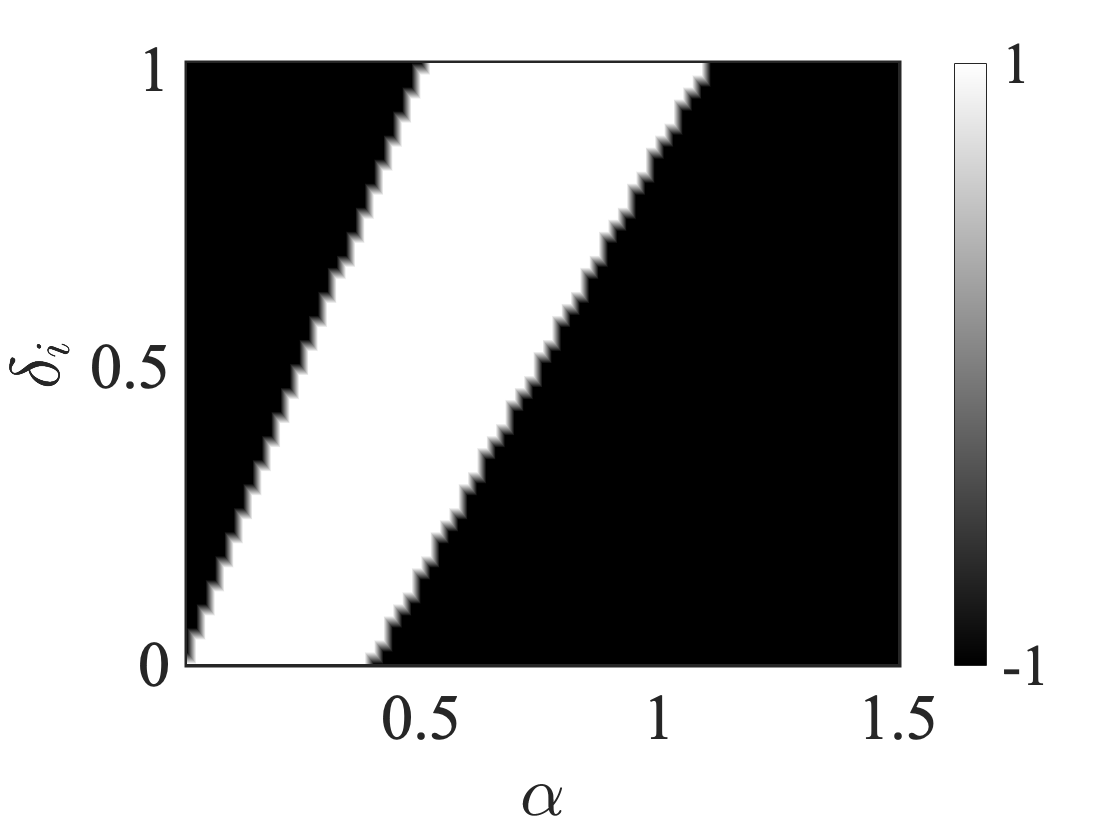}
    \caption{The sign of LMC with the inclusion of inelastic scattering. We choose $V^{\chi\chi'}_{\mathbf{k}\mathbf{k}'}\rightarrow 0$, $V^{\chi'\neq\chi}_{\mathbf{k}'\mathbf{k}}\rightarrow 0$, and thus $\alpha_i = V^{\chi\chi}_{\mathbf{k}'\mathbf{k}}/W^{\chi\chi'}_{\mathbf{k}\mathbf{k}'}$. Figure adapted from Ref.~\cite{sharma2023decoupling}.}
    \label{fig:inelastic}
\end{figure}

What is striking here is that even with weak internode scattering, LMC remains positive, contrasting with the scenario of strictly zero internode scattering (previous two subsections). Mathematically, this dissimilarity can be understood as follows: the distribution function obtained under weak internode scattering differs significantly from that in Fig.~\ref{fig:one_node_1}, where zero internode scattering was assumed. In the latter case, the distribution function's deviation must integrate to zero at a single node to conserve chiral charge. However, this requirement is relaxed in the presence of weak internode scattering, where only global charge conservation is necessary. A more physically transparent interpretation is presented next. 

The Boltzmann transport equation provides a steady-state solution applicable when $t$ greatly exceeds the maximum of $\tau_{\mathrm{inter}}$ and $\tau_{\mathrm{intra}}$. Thus, particles are free to scatter and redistribute across both nodes within a timescale much shorter than $\tau_\phi$. In order to pinpoint the role of $\tau_\phi$, it is incorporated it into the Boltzmann formalism~\cite{sharma2023decoupling}, and it is found that when $\tau_\mathrm{inter}\gg \tau_\phi$, one is in an effective one-node regime, and obtains negative LMC, which is are fully consistent with the results of a single-node obtained discussed earlier. The role of $\tau_\phi$ is discussed in detail in Sec.~\ref{sec_inelastic}.

These observations underscore several crucial points: (i) intranode scattering alone does not generate positive LMC due to chiral charge conservation, (ii) finite internode scattering is fundamental for observing positive LMC in the semiclassical low-$B$ limit, (iii) positive LMC induced by chiral anomaly is thus an exclusive internode phenomenon, bridging the Boltzmann and Landau-level perspectives, and (iv) surpassing a critical value $\alpha_c$, sufficiently large internode scattering switches the LMC sign from positive to negative. Consequently, $\tau_\phi\gg\tau_\mathrm{inter}>\tau_\mathrm{inter}^\mathrm{c}$ becomes imperative in experimental settings to observe positive LMC, where $\tau_\mathrm{inter}^\mathrm{c}$ represents the critical intervalley scattering time below which LMC turns negative. Fig.~\ref{fig:table} summarizes the conditions. 

\subsubsection{Inclusion of inelastic scattering} \label{sec_inelastic}
The discussion so far result in the following conundrum. Note that (i) LMC is negative when $\alpha=0$, (ii) LMC is positive when $\alpha>0$, (iii) LMC is negative when $\alpha>\alpha_c$. How does an infinitesimal intervalley scattering suddenly switch the sign of LMC? The inclusion of $\tau_\phi$, which is an inelastic scattering timescale, is imperative to resolve this issue and was presented in Ref.~\cite{sharma2023decoupling}. We briefly review this here.

\begin{figure}
    \centering
    \includegraphics[width=.99\columnwidth]
    {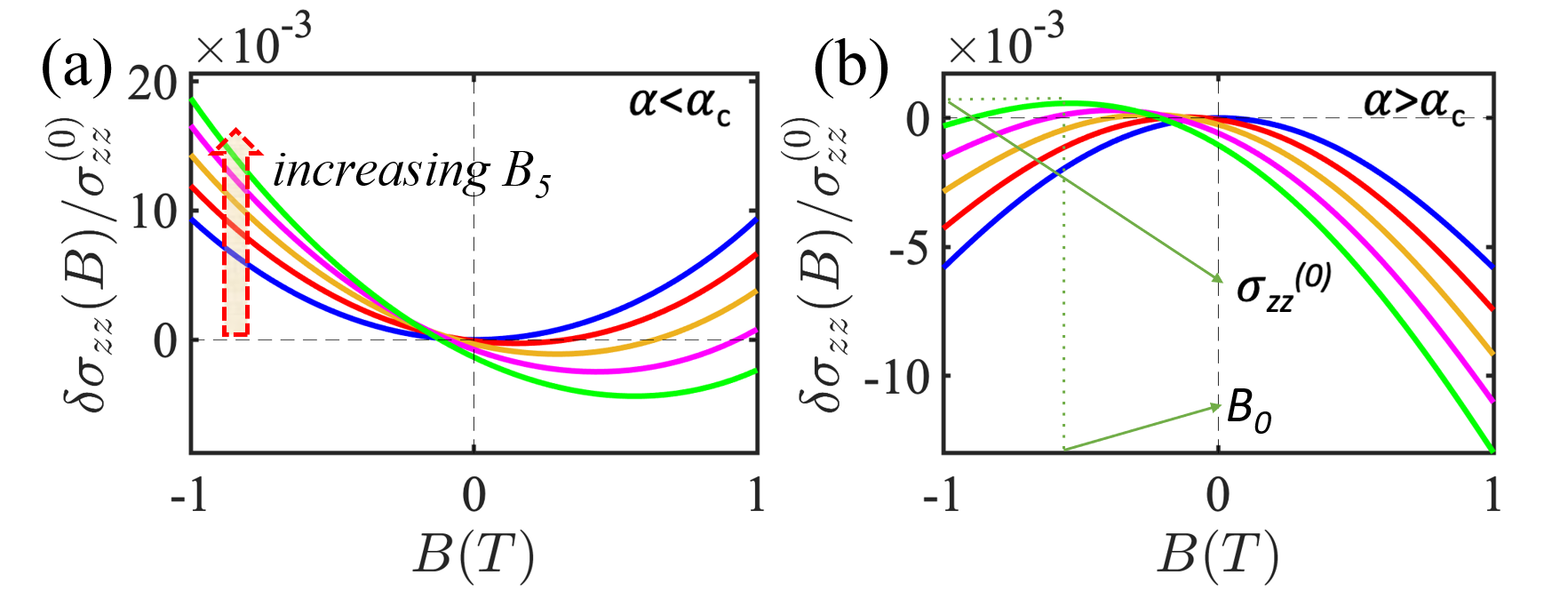}
    \caption{LMC v/s $\mathbf{B}$ at different values of $\mathbf{B_5}$ for the cases when intervalley scattering is less and greater than some critical value $\alpha_{c}$. For $\alpha<\alpha_{c}$ the effect of $\mathbf{B_{5}}$ field results in tilt and shift of the LMC parabola, which is dubbed as the week sign reversal, while for the case $\alpha>\alpha_{c}$ the LMC parabola is reversed, tilted and shifted, thus dubbed as \textit{strong} and \textit{weak} sign reversal. Figure adapted from Ref.~\cite{ahmad2023longitudinal}.}
\label{fig:LMC_vs_B_B5_vary}
\end{figure}

For elastic scattering, the scattering rate from state $|\mathbf{k}\rangle$ to $|\mathbf{k}'\rangle$ is identical to the rate from $|\mathbf{k}'\rangle$ to $|\mathbf{k}\rangle$. In other words, we can express the collision integral as:
\begin{align}
        \mathcal{I}_\mathrm{coll}[g^\chi_\mathbf{k}]=\sum\limits_{
    \chi'}\sum\limits_{\mathbf{k}'} \left(g^{\chi'}_{\mathbf{k}'} - g^{\chi}_{\mathbf{k}}\right) W^{\chi\chi'}_{\mathbf{k}\mathbf{k}'}.
\end{align}
We'll introduce another scattering rate, denoted as $V^{\chi\chi'}_{\mathbf{k}\mathbf{k}'}$, to encompass inelastic scattering within the medium. It follows that $V^{\chi\chi'}_{\mathbf{k}\mathbf{k}'} \neq V^{\chi\chi'}_{\mathbf{k}'\mathbf{k}}$, enabling us to express the total collision integral as:
\begin{align}
        \mathcal{I}_\mathrm{coll}[g^\chi_\mathbf{k}]
        &=\sum\limits_{
    \chi'}\sum\limits_{\mathbf{k}'} \left(g^{\chi'}_{\mathbf{k}'} - g^{\chi}_{\mathbf{k}}\right) W^{\chi\chi'}_{\mathbf{k}\mathbf{k}'}
    \nonumber\\ &+\sum\limits_{
    \chi'}\sum\limits_{\mathbf{k}'} \left(V^{\chi'\chi}_{\mathbf{k}'\mathbf{k}}g^{\chi'}_{\mathbf{k}'} - V^{\chi\chi'}_{\mathbf{k}\mathbf{k}'} g^{\chi}_{\mathbf{k}}\right) .
\end{align}
Substituting the ansatz for $g^\chi_\mathbf{k}$, the Boltzmann equation becomes 
\begin{align}
&\mathcal{D}^\chi \left[v^{\chi}_z + \frac{e B}{\hbar} (\boldsymbol{\Omega}^\chi\cdot \mathbf{v}^\chi_\mathbf{k})\right] = \sum\limits_{\chi'}\sum\limits_{\mathbf{k}'} W^{\chi\chi'}_{\mathbf{k}\mathbf{k}'} (\Lambda^{\chi'}_{\mathbf{k}'} - \Lambda^\chi_\mathbf{k}) \nonumber\\&+ \sum\limits_{\chi'}\sum\limits_{\mathbf{k}'} (V^{\chi'\chi}_{\mathbf{k}'\mathbf{k}} \Lambda^{\chi'}_{\mathbf{k}'} - V^{\chi\chi'}_{\mathbf{k}\mathbf{k}'}\Lambda^\chi_\mathbf{k}).
\end{align} 
The modified valley scattering rate is:
\begin{align}
\frac{1}{\tau^\chi_\mathbf{k}} = \mathcal{V} \sum\limits_{\chi'} \int{\frac{d^3 \mathbf{k}'}{(2\pi)^3} (\mathcal{D}^{\chi'}_{\mathbf{k}'})^{-1} \left(W^{\chi\chi'}_{\mathbf{k}\mathbf{k}'} + V^{\chi\chi'}_{\mathbf{k}\mathbf{k}'}\right)},
\label{Eq_tau13}
\end{align}
and the Boltzmann equation becomes:
\begin{align}
&\mathcal{D}^\chi \left[v^{\chi}_z + \frac{e B}{\hbar} (\boldsymbol{\Omega}^\chi\cdot \mathbf{v}^\chi_\mathbf{k})\right] +\frac{\Lambda^\chi_\mathbf{k}}{\tau^\chi_\mathbf{k}}\nonumber\\
&= \sum\limits_{\chi'}\sum\limits_{\mathbf{k}'} (W^{\chi'\chi}_{\mathbf{k}'\mathbf{k}} +V^{\chi'\chi}_{\mathbf{k}'\mathbf{k}})\Lambda^{\chi'}_{\mathbf{k}'} .
\end{align} 
To account for the impact of inelastic scattering on the valley scattering time, we can make the following substitution: $W^{\chi\chi'}_{\mathbf{k}\mathbf{k}'} \rightarrow - V^{\chi\chi'}_{\mathbf{k}\mathbf{k}'} + W^{\chi\chi'}_{\mathbf{k}\mathbf{k}'}$. With this modification, the valley scattering time maintains its original expression, while the Boltzmann equation transforms to:  
\begin{align}
\mathcal{D}^\chi \left[v^{\chi}_z + \frac{e B}{\hbar} (\boldsymbol{\Omega}^\chi\cdot \mathbf{v}^\chi_\mathbf{k})\right] +\frac{\Lambda^\chi_\mathbf{k}}{\tau^\chi_\mathbf{k}}= \sum\limits_{\chi'}\sum\limits_{\mathbf{k}'} W^{\chi'\chi}_{\mathbf{k}'\mathbf{k}} (1+\delta_i) \Lambda^{\chi'}_{\mathbf{k}'}, 
\end{align} 
where 
\begin{align}
    \delta_i = \frac{-V^{\chi\chi'}_{\mathbf{k}\mathbf{k}'}+V^{\chi'\chi}_{\mathbf{k}'\mathbf{k}}}{W^{\chi\chi'}_{\mathbf{k}\mathbf{k}'}}.
\end{align}
When $\delta_i=0$, the scenario corresponds to elastic scattering. Figure \ref{fig:inelastic} illustrates the sign of longitudinal magnetoconductance as a function of both the inelastic scattering coefficient $\delta_i$ and the intervalley scattering strength $\alpha$, both normalized with respect to the intravalley scattering strength.
First, considering the limit as $\delta_i\rightarrow 0$, implying $\tau_\phi$ as the largest timescale in the system. With $\tau_\mathrm{inter}$ large but still less than $\tau_\phi$ (i.e., when $\alpha\ll 1$), positive LMC is observed. Decreasing $\tau_\mathrm{inter}$ (increasing $\alpha$) leads to an inversion from positive to negative LMC at a critical value of intervalley scattering strength $\alpha_c$, consistent with previous discussion. 
Next, exploring scenarios with large but finite $\tau_\phi$ (i.e., $0<\delta_i<1$), we find that if $\alpha$ is smaller than $f\delta_i$, where $f$ represents a fraction less than one (i.e., when $\tau_\phi<f\tau_\mathrm{inter}$), one operates within the effective one-node regime, yielding negative LMC consistent with single-node results. The value of $f$, however, is non-universal and relies on microscopic parameters, with $f=1/2$ observed here.
Furthermore, by maintaining $\delta_i$ fixed, increasing $\alpha$ (i.e., decreasing $\tau_\mathrm{inter}$) results in an inversion of LMC sign at some critical value of the intervalley scattering strength.

\begin{figure}
    \centering
    \includegraphics[width=\columnwidth]{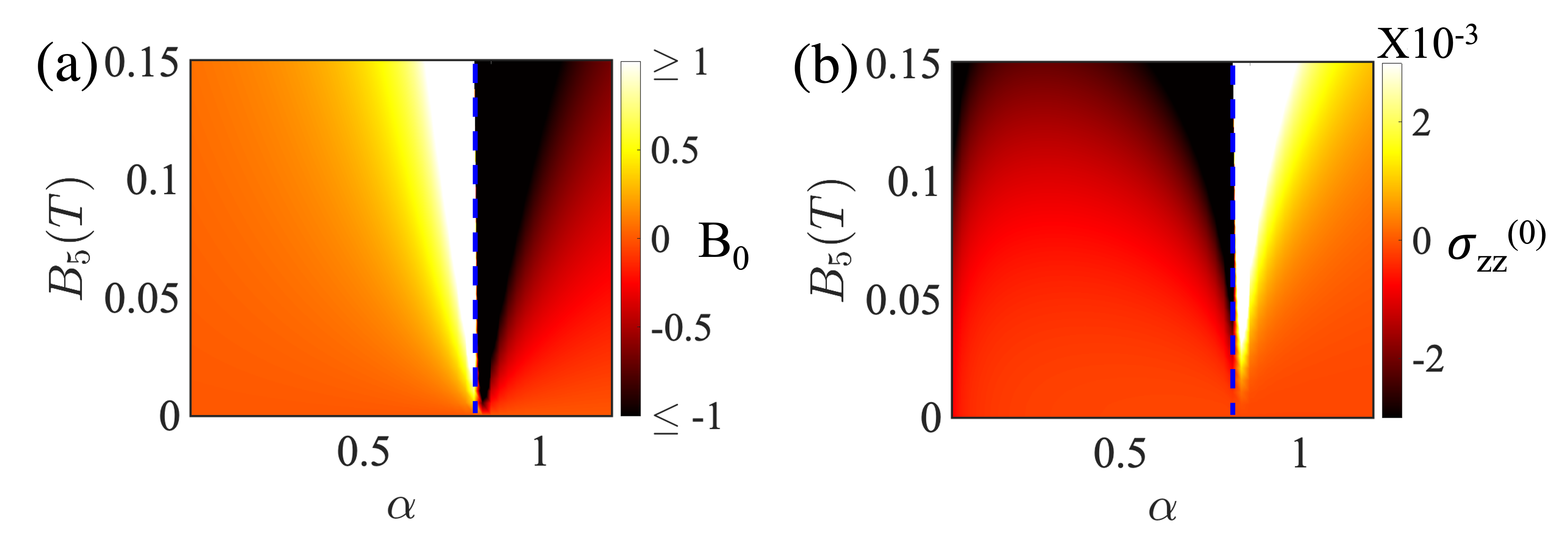}
    \caption{(a) The vertex of the parabola $B_0$, and (b) conductivity at $B_0$ for a minimal model of untilted TR broken WSM. Around the blue dashed contour ($\alpha=\alpha_c$) we see a `strong' sign-reversal. The parameters $B_0$ and $\sigma_{zz}^{(0)}$ show a striking change of sign as we move across the $\alpha_c$ contour. Figure adapted from Ref.~\cite{ahmad2023longitudinal}.}
    \label{fig:sigma002}
\end{figure}

\subsection{Magnetotransport in inhomogeneous Weyl semimetals}
\label{sec:inhomogeneous wsm}
Having discussed the nature of longitudinal magnetoconductance in weakly disordered Weyl semimetals, we now move to the discussion of magnetotransport in inhomogeneous WSMs and illustrate the effects of strain on magnetotransport. Recent studies~\cite{jackiw2007chiral,vozmediano2010gauge,guinea2010energy,cortijo2015elastic,pikulin2016chiral,grushin2016inhomogeneous,levy2010strain} have investigated the effect of elastic deformations (or strain) on massless Dirac fermions. Strain is shown to couple as an axial magnetic field (also termed a chiral gauge field) to Dirac/Weyl fermions.  In graphene, for instance, spectroscopic measurements of Landau levels reveal strain-induced fields reaching up to 300T~\cite{levy2010strain}. In Weyl semimetals, strain has been shown to produce adjustable variations in carrier mobility, electronic conductivity, and even phase transitions \cite{sisakht2016strain,niu2023high,thulin2008calculations,hong2008strain,yan2014effects,gui2008band,miao2022strain,rostami2015theory,peelaers2012effects,lazarovits2010effects,janotti2011strain,hwang1988effect,sakata2016effects,sabsovich2020pseudo,christensen1984electronic,guinea2010energy,arribi2020topological,jiang2015topological}.
The complex interactions between strain, electronic structure, and transport behavior are highlighted by several theoretical models and simulations, which shed light on the underlying mechanisms driving these phenomena \cite{ahmad2023longitudinal,gui2008band,ghosh2024electric,heidari2020chiral,yang2015chirality}.
\begin{figure*}
    \centering
    \includegraphics[width=2\columnwidth]{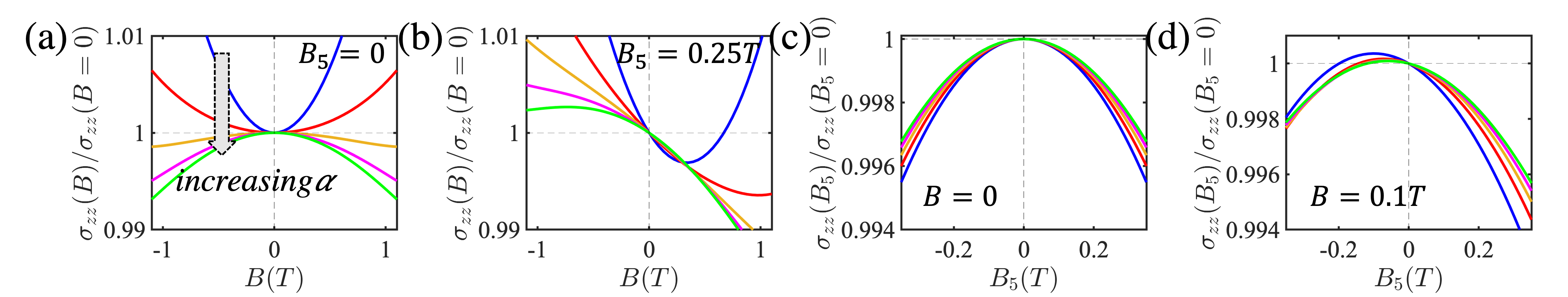}
    \caption{Longitudinal magnetoconductivity for a minimal model of TR broken untilted Weyl semimetal. (a) Increasing intervalley scattering strength results in strong sign reversal. (b) In addition to this, infinitesimal strain now results in weak sign-reversal as well.  (c) When plotted as a function of the gauge field $B_5$, LMC is always strongly sign-reversed. (d) In the presence of an external magnetic field, we see signatures of weak-sign reversal as well. 
    In all the plots as we move from blue to the green curve we increase the intervalley scattering strength $\alpha$ from below $\alpha_c$ to above $\alpha_c$. Figure adapted from Ref.~\cite{ahmad2023longitudinal}.}
    \label{fig:twonode_notilt_lmc_vs_B_vs_B5_vary_alpha}
\end{figure*}

In a minimal model of time-reversal symmetry broken Weyl semimetal, Weyl nodes are separated in momentum space by a vector $\mathbf{b}$ such that topological protection is ensured. The vector $\mathbf{b}$ is also interpreted as an axial gauge field and it couples with an opposite sign to Weyl nodes of opposite chirality \cite{goswami2013axionic,volovik1999induced,liu2013chiral,grushin2012consequences,zyuzin2012topological}. Spatially varying $\mathbf{b}$ thus generates an axial magnetic field $\mathbf{B}_5=\nabla\times \mathbf{b}$, which also couples oppositely to Weyl nodes of opposite chirality. An effective $\mathbf{B}_5$ field thus emerges from an inhomogeneous strain profile in Weyl semimetals. In the presence of this effective chiral gauge field $\mathbf{B}_5$, the effective magnetic field experienced by Weyl fermions at a given node of chirality $\chi$ is $\mathbf{B}\longrightarrow\mathbf{B}+\chi\mathbf{B}_5$. In a recent work Grushin \textit{et al.}~\cite{grushin2016inhomogeneous} have pointed out that even in the absence of an external magnetic field, the chiral gauge field influences the diffusive electron transport in Weyl semimetals by modifying its longitudinal magnetoconductance. Specifically, they show that strain, alone, results in positive LMC even in the absence of an external magnetic field. However, recent works, which consider momentum-dependent scattering beyond the relaxation-time approximation find contrary results~\cite{ahmad2023longitudinal,sharma2023decoupling}.
On similar lines, Ghosh~\textit{et al.}~\cite{ghosh2020chirality} point out that the planar Hall conductance can be manipulated by strain as well. However, contrary to this finding, refs.~\cite{ahmad2023longitudinal, sharma2023decoupling} conclude that the strain-induced contribution is not only opposite to that of the regular planar Hall effect but is also different in magnitude. We briefly review these findings below. 

Strain-induced magnetotransport can be studied using the semiclassical approximation discussed before by considering a valley-dependent magnetic field. First, we assume that $\mathbf{B}_5$ is held parallel to the electric $\mathbf{E}$ and the external magnetic field, $\mathbf{B}$, and calculate $\sigma_{zz}$. The behavior of $\delta\sigma_{zz}(B)$, or the change in LMC caused by the magnetic field, is plotted in Fig.~\ref{fig:LMC_vs_B_B5_vary}. This change in conductance can be expressed as follows: $\delta\sigma_{zz}(B) = \sigma_{zz}(B) - \sigma_{zz}(B=0)$. We see that the direction of the magnetic field, particularly near $B=0$, affects whether and how much LMC increases or decreases. The conductance increases for a negative magnetic field and decreases for a positive magnetic field. Moreover, the conductance increases (decreases) for both positive and negative values of $B$ when $\alpha<\alpha_c$ ($\alpha>\alpha_c$), when the magnitude of $B$ is increased further away from zero. We may compare this to the non-strained behavior, where the behavior of conductance depends on the strength of the magnetic field and either increases (when $\alpha<\alpha_c$) or decreases (when $\alpha>\alpha_c$) as seen in Fig.~\ref{fig:two_node_1}. 

In pursuit of clarifying the intricacies in magnetoconductivity behavior discussed above, Ahmad \textit{et al.}~\cite{ahmad2023longitudinal} begin by generalizing the magnetoconductivity expression as: 
\begin{align}
\sigma_{zz}(B)= \sigma_{zz}^{(2)}+ \sigma_{zz}^{(0)} + (B-B_0)^2.
\label{Eq-szz-fit}
\end{align}
To align with the observations depicted in Fig.~\ref{fig:LMC_vs_B_B5_vary}, the vertex of the parabola ($B_0$) is adjusted further from the origin utilizing the aforementioned definition. Consequently, while LMC is consistently positive around the vertex $B_0$ in Fig.~\ref{fig:LMC_vs_B_B5_vary}(a) when the strength of the intervalley scattering $\alpha<\alpha_c$, it is negative at small values of positive magnetic fields. In essence, LMC remains positive when considering the change in magnetic field and conductivity about the conductivity at $B_0$, rather than about the origin. The authors of Ref.~\cite{ahmad2023longitudinal} term this as \textit{weak} sign-reversal, as this phenomenon arises due to the preserved orientation of the parabola with only the vertex relocated away from the origin, and importantly $\sigma_{zz}^{(2)}$ retains its positive sign. Thus, strain in inhomogeneous Weyl semimetals induces the system into a state of \textit{``weak"} sign-reversal along a specific direction of the magnetic field when intervalley scattering is weak. Summarily, the features characterizing weak sign-reversal include (i) $B_0 \neq 0$, (ii) $\sigma_{zz}^{(0)} \neq \sigma_{zz}(B=0)$, and (iii) $\mathrm{sign }\; \sigma_{zz}^{(2)}>0$.

Conversely, in Fig.~\ref{fig:LMC_vs_B_B5_vary}(b), the orientation of the parabola is inverted upon surpassing the critical threshold of intervalley scattering strength ($\alpha_c$). Consequently, $\sigma_{zz}^{(2)}$ becomes negative, resulting in a decrease of LMC relative to $B_0$. This phenomenon is termed \textit{``strong"} sign-reversal~\cite{ahmad2023longitudinal}. Unlike its weak counterpart, strong sign-reversal lacks restrictions on the values of $B_0$ and $\sigma_{zz}^{(0)}$, being solely governed by the condition: (i) $\mathrm{sign }\; \sigma_{zz}^{(2)}<0$. In conclusion, the distinctive features of both \textit{``strong and weak"} sign-reversal are: (i) $B_0 \neq 0$, (ii) $\sigma_{zz}^{(0)} \neq \sigma_{zz}(B=0)$, and (iii) $\mathrm{sign }; \sigma_{zz}^{(2)}<0$.

In Fig.~\ref{fig:sigma002}, we show how the parameters $B_0$ and $\sigma_{zz}^{(0)}$  depend on the chiral gauge field and intervalley scattering strength. The transition between the `weak' and `strong and weak' cases (and vice versa) is characterized by an abrupt reversal in the signs of the relative offset in conductivity $\sigma_{zz}^{(0)}$, alongside the vertex of the parabola $B_0$, where $B_0\leq 0$ when $\sigma_{zz}^{(0)}\geq 0$, and vice versa. However, $\sigma_{zz}^{(2)}$ exhibits continuous interpolation across zero (not depicted in the plot). Notably, no discontinuity in $B_0$ or $\sigma^{(0)}$ is observed in the weak sign-reversed case. In other words, as the strain-induced field increases from zero for a constant intervalley scattering, the parameters $B_0$ and $\sigma_{zz}^{(0)}$ undergo continuous variation.

In Fig.\ref{fig:twonode_notilt_lmc_vs_B_vs_B5_vary_alpha}, the longitudinal magnetoconductivity is plotted as a function of magnetic field for various values of intervalley scattering. In the absence of a chiral gauge field (Fig.\ref{fig:twonode_notilt_lmc_vs_B_vs_B5_vary_alpha}(a)), as anticipated, we discern strong sign-reversal when $\alpha>\alpha_c$. However, in the presence of a chiral gauge field (Fig.~\ref{fig:twonode_notilt_lmc_vs_B_vs_B5_vary_alpha}(b)), we observe both strong and weak sign-reversal phenomena, as discussed before. In Fig.\ref{fig:twonode_notilt_lmc_vs_B_vs_B5_vary_alpha}(c), the longitudinal magnetoconductance is plotted exclusively as a function of the chiral gauge magnetic field (i.e., $B=0$). With no external magnetic field present in this scenario, the delineation of positive/negative LMC and weak/strong sign-reversal is solely being defined with respect to the $B_5$ field. It is seen that strain-induced chiral gauge field alone consistently yields a strong sign-reversed phase, irrespective of the intervalley scattering strength! Notably, this observation holds even in the presence of an external $B$-field (Fig.\ref{fig:twonode_notilt_lmc_vs_B_vs_B5_vary_alpha}(d)). 
Interestingly, earlier studies suggest that strain alone can lead to a positive contribution to the longitudinal magnetoconductance (LMC) even in the absence of an external magnetic field ($\mathbf{B}=0$)~\cite{grushin2016inhomogeneous}. However, Fig.~\ref{fig:twonode_notilt_lmc_vs_B_vs_B5_vary_alpha} reveals a striking contrast: strain alone results in a decrease in conductance. 
In the absence of strain, the transition of LMC from positive to negative is also attributed to the effect of the orbital magnetic moment (OMM). However, even when disregarding the contribution of OMM, the results shown in Fig.~\ref{fig:twonode_notilt_lmc_vs_B_vs_B5_vary_alpha} remain qualitatively unchanged. This is because, unlike in the former scenario, strain-induced OMM impacts both nodes equally, thereby preserving the similarity of the Fermi surfaces.

\subsection{Planar Hall effect}
\label{sec:phe}
We now discuss another aspect of the chiral anomaly, specifically the planar Hall effect (PHE)~\cite{nandy2017chiral, burkov2017}. This effect occurs when there is an in-plane transverse voltage due to the misalignment of coplanar electric and magnetic fields. The planar Hall conductivity (PHC), denoted as $\sigma_{zx}$, measures the transverse conductivity along the $\hat{z}$ direction, perpendicular to both the applied electric field and the current in the $\hat{z}$ direction, and under the influence of a magnetic field in the $x$-$z$ plane that makes an angle $\gamma$ with respect to the $x$-axis (see Fig.~\ref{fig:WSM_slab_nd_Landau_picture}). This phenomenon is well-known in ferromagnetic systems~\cite{vu1968,ge2007,friedland2006, goennenwein2007,bowen2005}, showing an angular dependence similar to that observed in Weyl semimetals (WSMs). In Ref.~\cite{nandy2017chiral}, Nandy \textit{et al.} developed a semiclassical transport theory of the PHE linking it to the occurrence of chiral anomaly and Berry curvature. Subsequent works~\cite{ma2019planar,medel2023planar,das2023chiral,kumar2018planar,yang2019current,li2018giant,chen2018planar,li2018giant2,pavlosiuk2019negative,singha2018planar,sharma2019transverse} have studied PHE and its corresponding thermal transport responses in the context of Weyl, Dirac and spin-orbit coupled metals, using it as another probe of the chiral anomaly in these materials.
The crucial role of internode scattering in the planar Hall effect considering momentum-dependent scattering and charge conservation was studied in Ref.~\cite{sharma2020sign} where it was concluded that unlike LMC, the planar Hall conductivity does not switch sign for any value of the internode scattering strength. Furthermore, the behavior with respect to the external magnetic field is quadratic, and the dependence on the orientation of the magnetic field is $\sin (2\gamma)$ irrespective of the strength of intervalley scattering.

\begin{figure}
    \centering
    \includegraphics[width=\columnwidth]{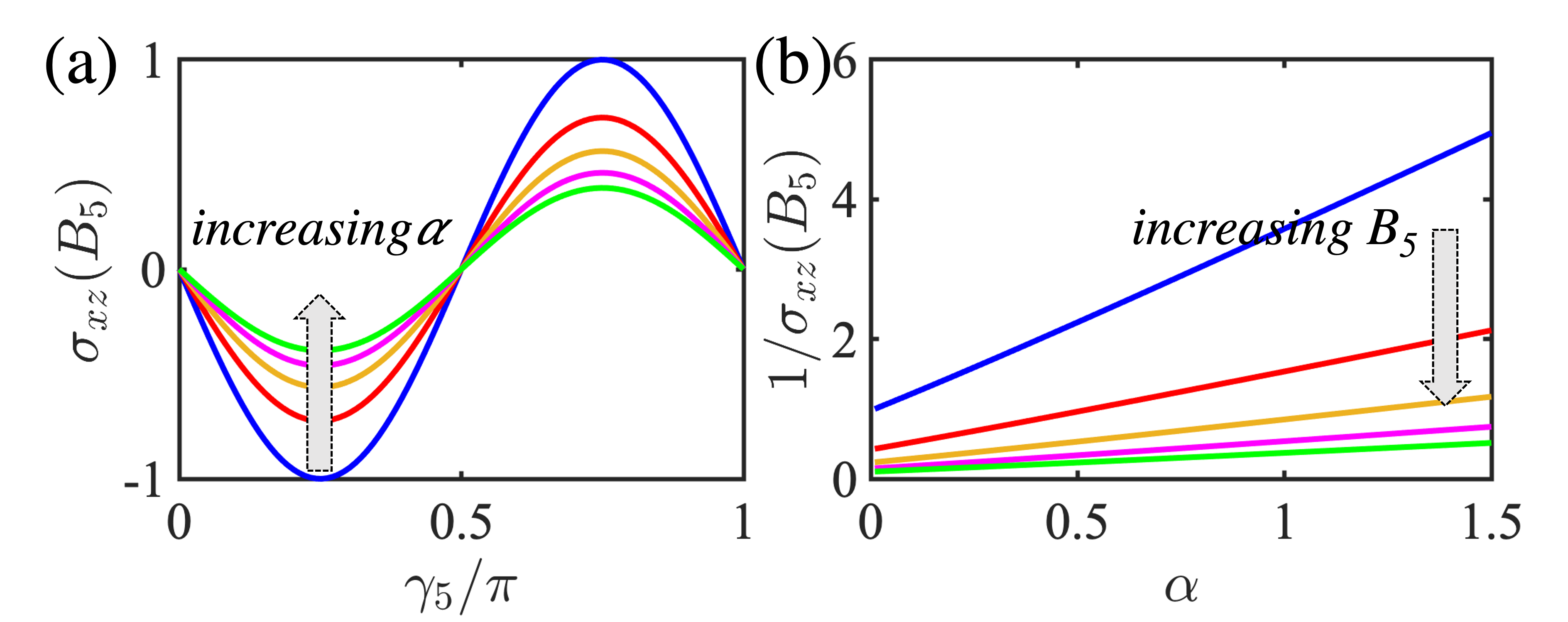}
    \caption{Planar Hall conductivity for a minimal model of untilted TR broken WSM in the absence of any magnetic field. (a) Variation with respect to the angle $\gamma_5$. Increasing $\alpha$ reduces the conductivity. (b) PHC behaves as the inverse of scattering strength. Since $\sigma_{xz}(B_5=0)=0$, we have normalized $\sigma_{xz}$ appropriately in both the plots. Increasing the $B_5$ field increases the conductivity. Figure adapted from Ref.~\cite{ahmad2023longitudinal}.}
    \label{fig:phc001}
\end{figure}

Similar to LMC, the dependence on the magnetic field is typically quadratic and we may expand the planar Hall conductivity $\sigma_{xz}$ as
\begin{align}
\sigma_{xz}(B)= \sigma_{xz}^{(2)}(B-B_0)^2 + \sigma_{xz}^{(0)},
\label{eq:phc1}
\end{align} 
where $B_0$ is vertex of the parabola, and $\sigma_{xz}^{(2)}$ is the quadratic coefficient. To discuss the effect of strain, we begin by evaluating the planar Hall conductivity in the absence of an external magnetic field~\cite{ahmad2023longitudinal}. Fig.~\ref{fig:phc001} presents the planar Hall conductivity $\sigma_{xz}(B_5)$, computed without an external magnetic field. The angular dependence on $\gamma_5$ follows the relation $\sim \sin (2\gamma_5)$, consistent with the behavior observed in conventional planar Hall conductivity~\cite{ahmad2023longitudinal}. Additionally, Ahmad \textit{et al.} examined the impact of intervalley scattering~\cite{ahmad2023longitudinal}, and their numerical analysis reveals that the planar Hall conductivity induced by the chiral gauge field behaves as $\sim 1/\alpha$. 

\begin{figure}
    \centering
    \includegraphics[width=\columnwidth]{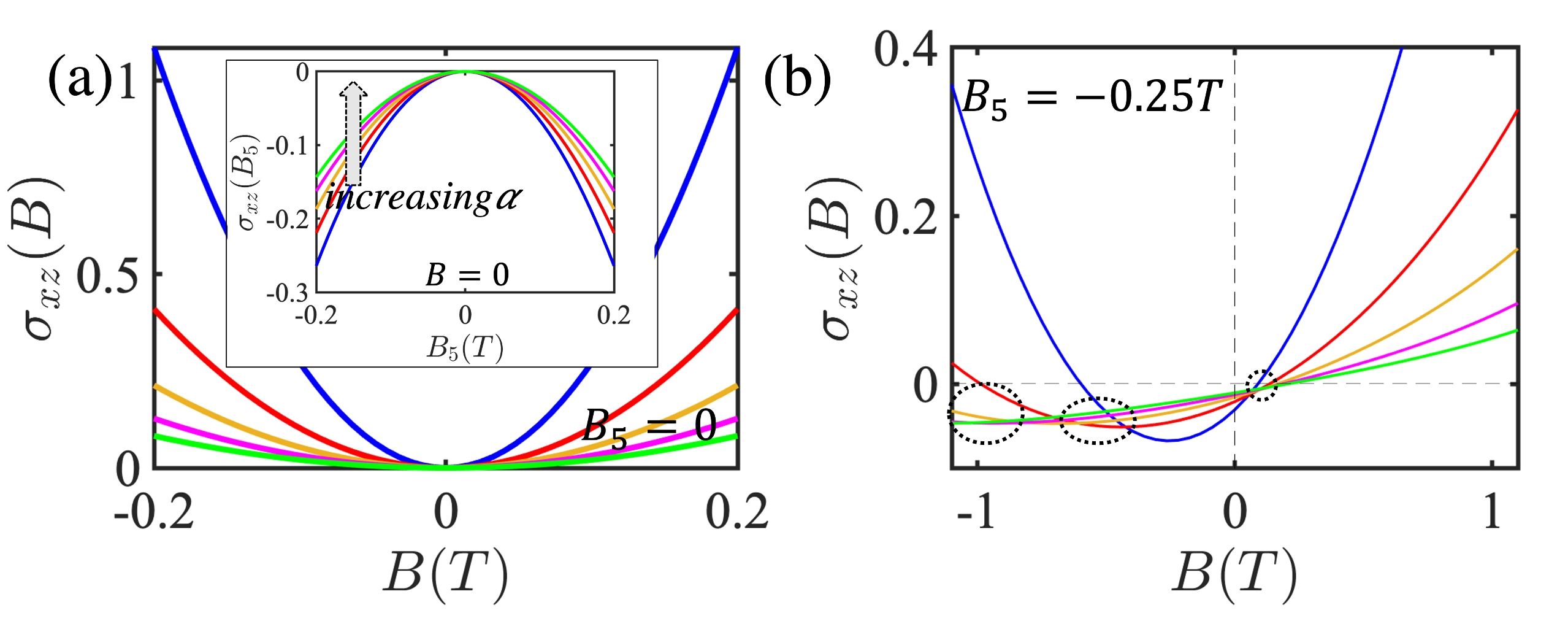}
    \caption{Planar Hall conductivity for a minimal model of untilted TR broken WSM. (a) PHC as a function of the external magnetic field $B$ and no strain-induced field ($B_5=0$). Inset: PHC has been plotted as a function of $B_5$ with no external field ($B=0$). The angle $\gamma$ was chosen to be equal to $\gamma_5$. (b) PHC in the presence of both magnetic field and strain. The chiral gauge field causes weak sign-reversal. The dotted ellipses highlight regions that show an anomalous behavior with respect to intervalley scattering strength.  All the plots are appropriately normalized. Figure adapted from Ref.~\cite{ahmad2023longitudinal}.}
    \label{fig:phc002}
\end{figure}

\begin{figure}
    \centering
    \includegraphics[width=\columnwidth]{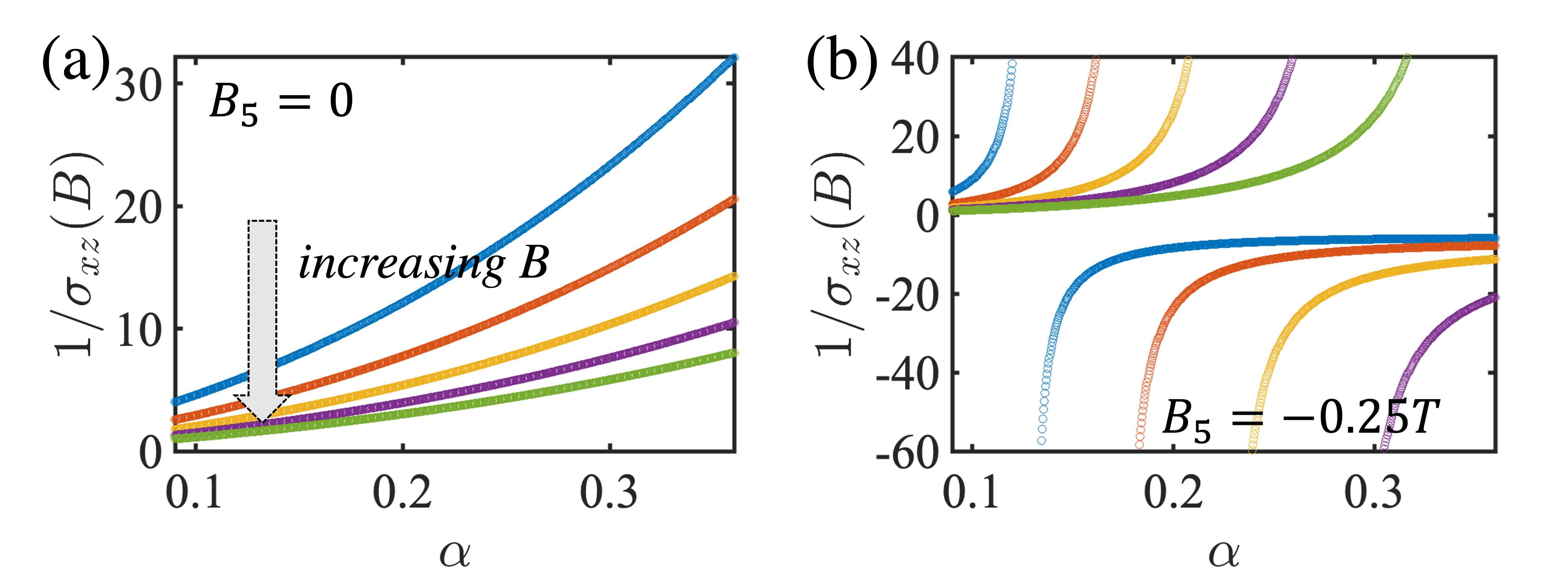}
    \caption{Planar Hall conductance for a minimal model of untilted WSM as a function of intervalley scattering strength. (a) in absence of $B_5$ field. (b) in presence of $B_5$ field. In all the curves, as we go from blue to green, we increase $B$. All the plots are appropriately normalized. Figure adapted from Ref.~\cite{ahmad2023longitudinal}.}
    \label{fig:twonode_notilt_inv_sxz}
\end{figure} 

\begin{figure*}
    \centering
    \includegraphics[width=2\columnwidth]{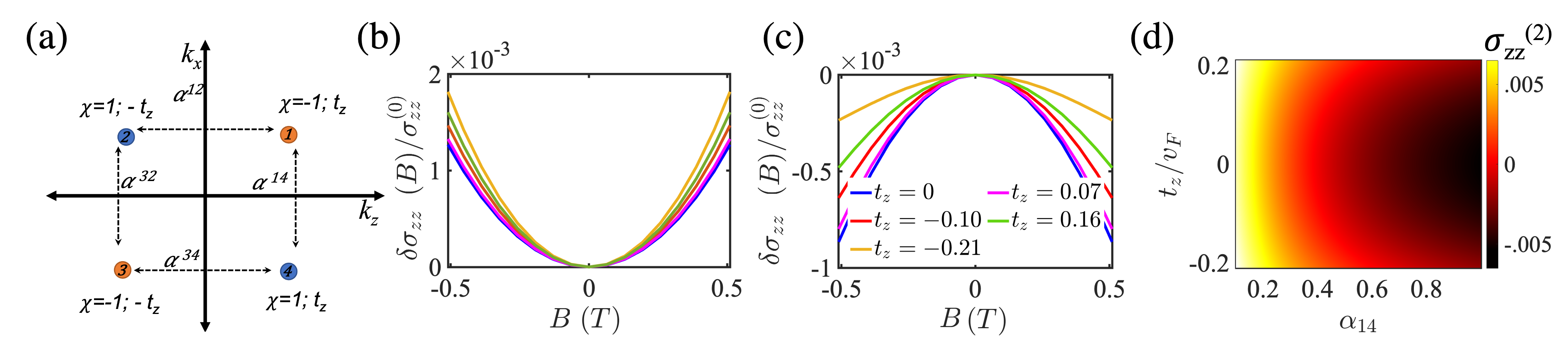}
    \caption{(a) Schematic of Weyl nodes in a prototype model of an inversion asymmetric Weyl semimetal. Here $\chi$ is the chirality, $t_z$ is the tilt, and $\alpha^{ij}$ are scattering rates from node $i$ to node $j$. (b) LMC as a function of magnetic field when the intervalley scattering rates are less than the critical value. (c) LMC as a function of magnetic field when the intervalley scattering rates are above the critical value. The legends in (b) and (c) are identical. (d) $\sigma_{zz}^{(2)}$ for a fixed value of $\alpha_{12}=0.19$. Plots (b), (c), and (d) are in the absence of strain, i.e., $B_5=0$. Figure adapted from Ref.~\cite{ahmad2023longitudinal}.}
    \label{fig:fournodelmc1}
\end{figure*}
\begin{figure*}
    \centering
    \includegraphics[width=2\columnwidth]{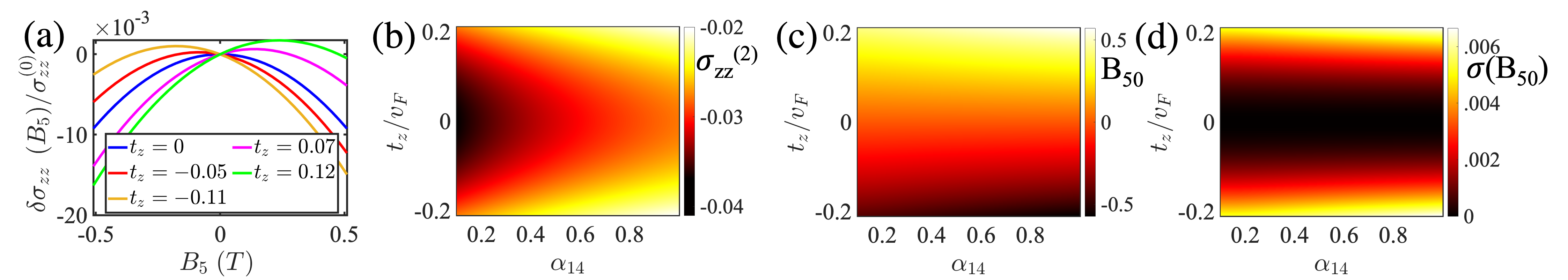}
    \caption{LMC for inversion asymmetric Weyl semimetal in the presence of strain induced chiral magnetic field ($B_5$) and absence of magnetic field. (a) A finite tilt can result in weak sign-reversal. The plot is for a fixed value of $\alpha_{12}=0.4$, but the qualitative behavior is independent of scattering strength. (b), (c), and (d) plot the parameters $\sigma_{zz}^{(2)}$, $B_{50}$, and $\sigma(B_{50})$ as a function of parameters $\alpha_{14}$ and $t_z$. We fixed $\alpha_{12}=0.19$. Figure adapted from Ref.~\cite{ahmad2023longitudinal}.}
    \label{fig:fournodelmc2}
\end{figure*}
\begin{figure}
    \centering
    \includegraphics[width=\columnwidth]{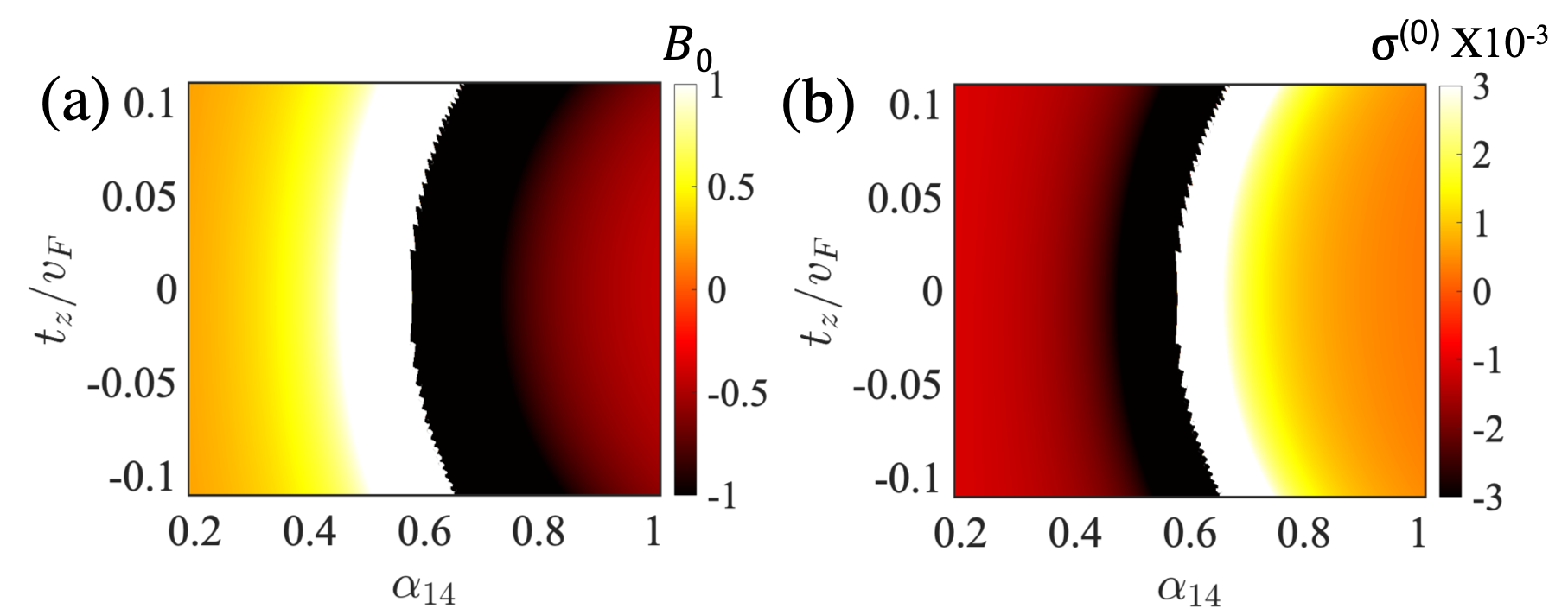}
    \caption{The parameters $B_0$ (a) and $\sigma^{(0)}$ (b) for inversion asymmetric Weyl semimetals (Eq.~\ref{eq_H4nodes}) {in the presence of both strain-induced field and external magnetic field}. We have fixed $\alpha_{12}=0.3$, $B_5=0.1T$. Weak sign reversal is not observed and strong sign-reversal occurs at $\alpha_{14}=\alpha_{14c}(t_z)$. Figure adapted from Ref.~\cite{ahmad2023longitudinal}.}
    \label{fig:fournodes_lmc_colorplot_B0_sigatB0_1}
\end{figure}
\begin{figure*}
    \centering
    \includegraphics[width=2\columnwidth]{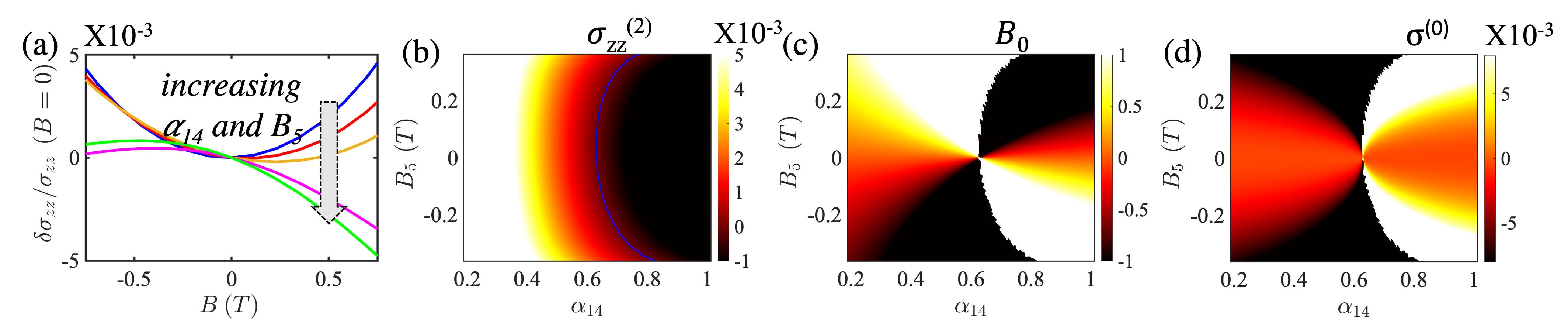}
    \caption{(a) LMC for inversion asymmetric Weyl semimetal. As we move from the blue to the green curve, we simultaneously increase $B_5$ as well as $\alpha_{14}$. Both weak and strong sign-reversal are exhibited. The plots (b), (c), and (d) plot the parameters $\delta\sigma_{zz}^{(2)}$, $B_0$, and $\sigma^{(0)}$ for fixed $\alpha_{12}$ and $t_z\neq 0$. The blue contour in plot (b) separates the phases where $\sigma_{zz}^{(2)}$ changes sign. Again, we see signatures of both weak and strong sign-reversal. The tilt parameter is fixed to $t_z/v_F=-0.1$. Figure adapted from Ref.~\cite{ahmad2023longitudinal}.}
    \label{fig:fournodes_lmc_colorplot_varyB5_vary_a14}
\end{figure*}

\begin{figure}
    \centering
    \includegraphics[width=\columnwidth]{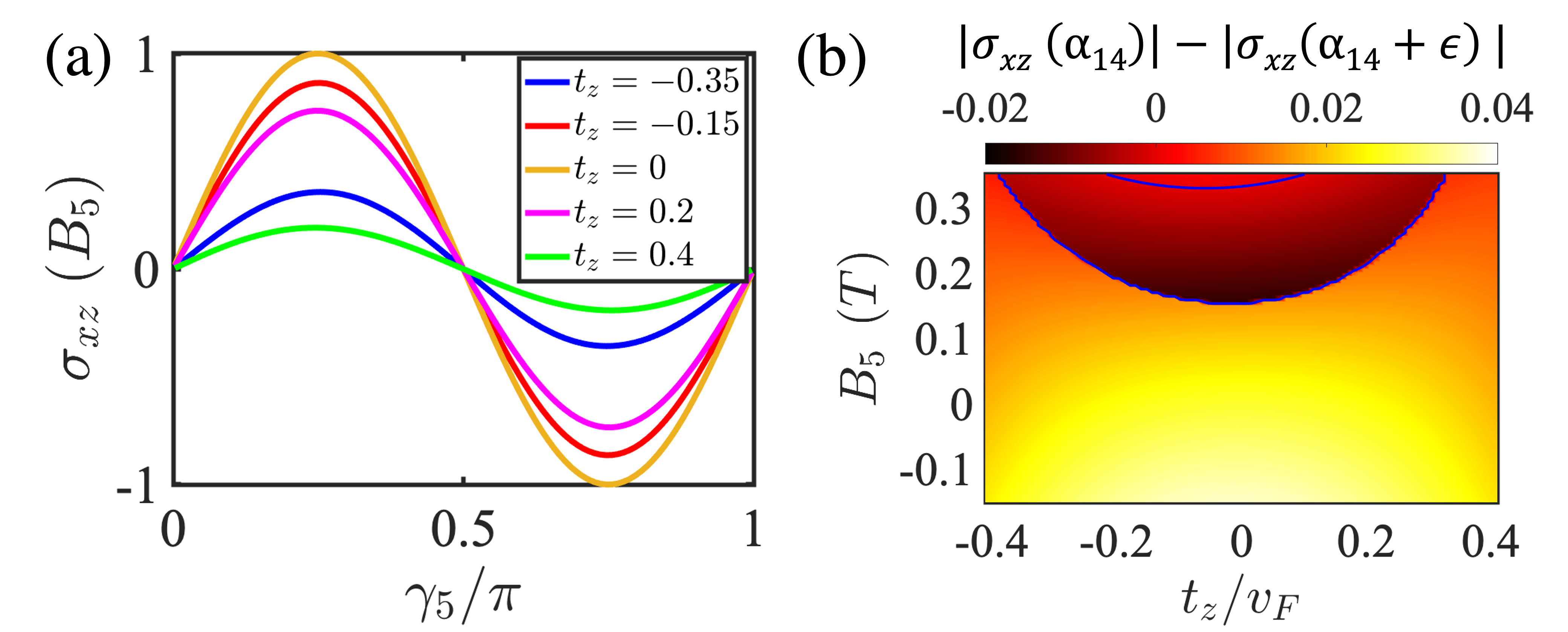}
    \caption{Planar Hall conductance for inversion asymmetric Weyl semimetal. (a) PHC as a function of $\gamma_5$, when $B=0$, and $B_5\neq 0$. (b) The change in the magnitude of the planar Hall conductivity on increasing $\alpha_{14}$ infinitesimally. In the region between the blue contours, we observe an anomalous increase in conductivity. Here we fix, $B=1T$, $\alpha_{12}=0.4$, $\alpha_{14}=0.5$, and $\epsilon=0.01$. Figure adapted from Ref.~\cite{ahmad2023longitudinal}.}
    \label{fig:fournodes_sxz_colorplot}
\end{figure}

In Ref.~\cite{ahmad2023longitudinal}, the authors investigate the behavior of the planar Hall conductivity under two distinct conditions: (a) when an external magnetic field is applied and the strain-induced field is absent, and (b) when the strain-induced field is present and the external magnetic field was absent. It is found that the contributions to the planar Hall conductivity differ both in sign and magnitude,  contrasting earlier claims~\cite{ghosh2020chirality}. Specifically, $\sigma_{xz}(B)$ increases with increasing $B$, while $\sigma_{xz}(B_5)$ decreases with increasing $B_5$. This feature is illustrated in Fig.~\ref{fig:phc002} (a). In other words, the chiral gauge field alone results in a strong sign-reversal. This has been attributed to the inclusion of intervalley scattering, momentum dependence, and charge conservation, which were neglected in previous studies.

When both external magnetic field and chiral gauge fields are present, the strain effect shifts and tilts the conductivity parabola, leading to a weak sign-reversal of the conductivity, as illustrated in Fig.~\ref{fig:phc002}(b). Unlike longitudinal magnetoconductivity, the planar Hall conductivity (PHC) does not exhibit a strong sign-reversal, even when the intervalley scattering exceeds the critical value.
Interestingly, within a specific window of the magnetic field, increasing the intervalley scattering strength enhances the magnitude of the planar Hall conductivity, which seems counter-intuitive. This behavior can be understood by considering the opposing effects of strain-induced PHC and magnetic field-induced PHC, each contributing oppositely and unequally to the overall planar Hall conductivity~\cite{ahmad2023longitudinal}.
This phenomenon is more clearly visualized in Fig.~\ref{fig:twonode_notilt_inv_sxz}, where the planar Hall conductivity is plotted as a function of the intervalley scattering strength $\alpha$. In the absence of a $B_5$ field, the Hall conductivity is non-linear as a function of $1/\alpha$, which is in contrast to Fig.~\ref{fig:phc001}(b) (where $B=0$, $B_5\neq 0$), which exhibits linear behavior across all ranges of $\alpha$. Furthermore, in the presence of a $B_5$ field, the behavior of $\sigma_{xz}$ with respect to $\alpha$ can be markedly different. Due to the weak sign-reversal, $\sigma_{xz}$ can switch its sign, explaining the divergences observed in Fig.~\ref{fig:twonode_notilt_inv_sxz}(b). Additionally, when $\sigma_{xz}$ transitions from positive to negative, its dependence on $\alpha$ becomes anomalous; specifically, increasing $\alpha$ increases the magnitude of $\sigma_{xz}$. Such an anomalous behavior with respect to the intervalley scattering strength is not observed in longitudinal magnetoconductivity.

\subsection{Inversion asymmetric WSM}
Having discussed the model of a simple time-reversal symmetry broken Weyl semimetal, here we review inversion asymmetric WSMs. Following Ref.~\cite{ahmad2023longitudinal}, we focus on a minimal model for an inversion asymmetric WSM consisting of four nodes. The Hamiltonian for this system is given by:
\begin{align}
H = \sum\limits_{n=1}^4 \left(\chi_{n}\hbar v_F \mathbf{k}\cdot\boldsymbol{\sigma} + \hbar v_F t_z^n k_z \sigma_0\right),
\label{eq_H4nodes}
\end{align}
where the system comprises four Weyl nodes located at $\mathbf{K} = (\pm k_0, 0, \pm k_0)$ in the Brillouin zone. In Eq.~\eqref{eq_H4nodes}, $\chi_n$ denotes the chirality, and $t_z^n$ represents the tilting of the Weyl cone, which is assumed to be along the $z$ direction. Specifically, we have:
\begin{align}
(1, t_z) &= (\chi_1, t_z^{(1)}) = (-\chi_2, t_z^{(2)}) \\
&= (\chi_3, -t_z^{(3)}) = (-\chi_4, -t_z^{(4)}),
\end{align}
such that inversion symmetry is broken. The tilt parameter $t_z$ is considered to be less than unity. Fig.~\ref{fig:fournodelmc1}(a) illustrates this prototype inversion asymmetric Weyl semimetal.
There are four intranode scattering channels (node $n \leftrightarrow n$) and four internode scattering channels (node $n \leftrightarrow [n+1] \mod 4$). The dimensionless scattering strength between nodes $m$ and $n$ is denoted as $\alpha^{mn}$. For simplicity, scattering between nodes (4 $\leftrightarrow$ 2) and (1 $\leftrightarrow$ 3) is ignored. The internode scatterings are categorized as follows:
\begin{enumerate}
\item Scattering between Weyl cones of opposite chirality and opposite tilt orientation (1 $\leftrightarrow$ 2 and 3 $\leftrightarrow$ 4).
\item Scattering between Weyl cones of opposite chirality and same tilt orientation (1 $\leftrightarrow$ 4 and 2 $\leftrightarrow$ 3).
\end{enumerate}
These two categories exhibit different behaviors, and uncovering their interplay is of particular interest.

We first explore the behavior of longitudinal magnetoconductivity in the absence of strain. Previously, it was established that for a system with two tilted cones of opposite chirality, `weak' sign-reversal is possible only if the cones are oppositely oriented. In the current scenario, however, the `weak' sign-reversal produced by the internode scattering channel (1 $\Longleftrightarrow$ 2) is exactly offset by the scattering channel (4 $\Longleftrightarrow$ 3). Furthermore, the scattering channels (1 $\Longleftrightarrow$ 4) and (2 $\Longleftrightarrow$ 3) do not induce weak sign-reversal because they involve Weyl cones with the same tilt. Thus, in the absence of a $B_5$ field, weak sign-reversal is not observed in the case of an inversion asymmetric WSM.

In Fig.~\ref{fig:fournodelmc1}, the longitudinal magnetoconductivity for the inversion asymmetric Weyl semimetal described by Eq.~\ref{eq_H4nodes} is plotted without the strain-induced chiral gauge field $B_5$. As discussed, no weak sign-reversal is observed; only strong sign-reversal occurs when $\alpha_{12}$ and/or $\alpha_{14}$ are sufficiently large. Increasing the tilt does not qualitatively alter this behavior; rather, an increase in the magnitude of the tilt in either direction only enhances the magnetoconductivity.

Next, we review the behavior in the absence of an external magnetic field but in the presence of a strain-induced gauge field $B_5$. Similar to time-reversal broken Weyl semimetals, the strain-induced chiral magnetic field $B_5$ results in a negative longitudinal magnetoconductivity (LMC) coefficient $\sigma_{zz}^{(2)}$. This finding contradicts previous claims that strain increases longitudinal magnetoconductivity~\cite{grushin2012consequences,ghosh2020chirality}. The contradiction arises due to the omission of intervalley scattering, momentum-dependent scattering, and charge conservation in these earlier studies, which are all considered in Ref.~\cite{ahmad2023longitudinal}.
Furthermore, it is revealed that strain alone induces strong sign-reversal, while the presence of tilt additionally causes weak sign-reversal. Fig.~\ref{fig:fournodelmc2}(a) illustrates LMC as a function of the strain-induced magnetic field $B_5$, showcasing these features.
The magnetoconductivity is modeled by the following expression:
\begin{align}
\sigma_{zz}(B_5) = \sigma_{zz}^{(2)}(B - B_{50})^2 + \sigma_{zz}(B_{50}).
\label{eq:sigma_25}
\end{align}
The slope $\sigma_{zz}^{(2)}$ is consistently negative, irrespective of tilt, strain, or intervalley scattering strengths across the nodes. The parabola's center, $B_{50}$, correlates directly with the tilt parameter $t_z$. The sign of $t_z$ determines whether $B_{50}$ is positive or negative. Although $B_{50}$ also depends on the scattering strength, this dependence is weaker compared to its dependence on $t_z$.
Fig.~\ref{fig:fournodelmc2}(b), (c), and (d) plot the parameters $\sigma_{zz}^{(2)}$, $B_{50}$, and $\sigma_{zz}(B_{50})$ as functions of $\alpha_{14}$ and $t_z$, while keeping $\alpha_{12}$ fixed and $B = 0$. No sharp discontinuities are observed in these parameters since there is already a strongly sign-reversed state.

In inversion asymmetric inhomogeneous Weyl semimetals, the interplay between the strain-induced chiral gauge field, external magnetic field, and the tilt parameter produces intriguing effects, as demonstrated in Fig.~\ref{fig:fournodes_lmc_colorplot_B0_sigatB0_1}. The longitudinal magnetoconductivity (LMC) is analyzed as a function of the external magnetic field for a fixed chiral gauge field. Eq.~\ref{Eq-szz-fit} is used to determine the fit parameters $B_0$, $\sigma_{zz}^{(2)}$, and $\sigma_{zz}^{(0)}$.
These results show that weak sign-reversal is absent. Instead, strong sign-reversal appears when the intervalley scattering parameter $\alpha_{14}$ exceeds a critical value $\alpha_{14c}$, which depends on the tilt parameter. Near the critical value $\alpha_{14c}(t_z)$, there is a pronounced shift in the signs of the parameters $B_0$ and $\sigma_{zz}^{(0)}$, indicating a continuous change in the sign of $\sigma_{zz}^{(2)}$. This behavior highlights the complex interaction between strain, magnetic field, and tilt in these materials.
Identifying the parameters $B_0$ and $\sigma_{zz}^{(0)}$ from experimentally measured conductivity can reveal the dominant scattering mechanisms in the system, such as internode or intranode scattering. These parameters also provide insights into the strain present in the samples and the tilting of the Weyl cones. Experimentally, LMC in inversion asymmetric Weyl semimetals can be explored by adjusting the strain in the system, making it crucial to study the effects of varying strain on LMC.

Fig.~\ref{fig:fournodes_lmc_colorplot_varyB5_vary_a14}(a) shows $\delta\sigma_{zz} = \sigma_{zz}(B) - \sigma_{zz}(B=0)$ as both the intervalley scattering strength $\alpha_{14}$ and the strain-induced chiral gauge field $B_5$ are varied. The plot reveals the presence of both weak and strong sign-reversal effects. When $\alpha_{14}$ exceeds the critical value $\alpha_{14c}$, strong sign-reversal occurs, while variations in the tilt parameter lead to weak sign-reversal.
By keeping $\alpha_{12}$ fixed, we can fit the parameters of $\sigma_{zz}(B)$ using Eq.~\ref{Eq-szz-fit}. Fig.~\ref{fig:fournodes_lmc_colorplot_varyB5_vary_a14}(b) displays $\sigma_{zz}^{(2)}$ as a function of $B_5$ and $\alpha_{14}$. The critical contour $\alpha_{14c}$, where $\sigma_{zz}^{(2)}$ changes sign, is influenced by $B_5$, indicating that $\alpha_{c}$ is generally a function of both the tilt parameter $t_z$ and $B_5$.
Figs.~\ref{fig:fournodes_lmc_colorplot_varyB5_vary_a14}(c) and (d) plot $B_0$ and $\sigma_{zz}^{(0)}$, respectively, derived from Eq.~\ref{Eq-szz-fit}. These plots exhibit interesting behavior as $B_5$ and $\alpha_{14}$ are varied. In Fig.~\ref{fig:fournodes_lmc_colorplot_varyB5_vary_a14}(c), when $\alpha < \alpha_c(B_5)$, $B_0$ shifts from negative to positive as $B_5$ transitions from negative to positive. When $\alpha > \alpha_c(B_5)$, $B_0$ changes from positive to negative. At the critical point $\alpha = \alpha_c(B_5)$, there is a pronounced sign-reversal, creating sharp contrasts on either side of $\alpha_c(B_5)$.
In Fig.~\ref{fig:fournodes_lmc_colorplot_varyB5_vary_a14}(d), $\sigma_{zz}^{(0)}$ does not change sign with variations in $B_5$, but it shows significant behavior around $\alpha_{c}(B_5)$ due to strong sign-reversal, similar to $B_0$. This behavior underscores the complex interplay between strain, scattering, and tilt in influencing LMC in these materials.

We also review the planar Hall effect in inversion asymmetric Weyl semimetals. Fig.~\ref{fig:fournodes_sxz_colorplot}(a) presents the planar Hall conductivity (PHC) $\sigma_{xz}$ as a function of the angle $\gamma_5$ when there is no external magnetic field but a strain-induced gauge field $B_5$ is present. The PHC follows a $\sim \sin(2\gamma_5)$ pattern. This occurs because the contributions from pairs of Weyl nodes that are time-reversed and have opposite tilts add together, while the contributions from pairs of Weyl nodes that are time-reversed and have the same tilt cancel out. This cancellation is why we do not observe a $\sim \sin(\gamma_5)$ trend.
In Fig.~\ref{fig:fournodes_sxz_colorplot}(b), we plot the change in the magnitude of the planar Hall conductivity when the intervalley scattering strength $\alpha_{14}$ is slightly increased. The plot indicates a region in the $B_5-t_z$ space where the conductivity change is unusual—increasing the intervalley scattering leads to an increase in the conductivity's magnitude. This behavior is consistent even when $\alpha_{14}$ is held constant and $\alpha_{12}$ is varied, hence it is not shown separately.
These findings underscore the nuanced interactions between strain-induced fields, scattering strengths, and the tilting of Weyl cones, which collectively influence the planar Hall effect in inversion asymmetric Weyl semimetals.
\section{Outlook}
The problem of electron flow in materials has always garnered the attention of physicists. Despite the complex nature of the underlying many-body quantum system, the Boltzmann theory reasonably captures the essential physics in metallic systems, especially when effects originating from quantum interference and entanglement are unimportant. This remains true as long as Bloch's theorem remains a valid assumption in the system. Matter, which was classified by Landau's paradigm of broken symmetry has now evolved into topological matter, where fundamental ideas from topology and geometry have allowed us to study physical systems from a new lens. 
It is remarkable that these ideas of quantum geometry and topology can be incorporated in the Bloch-Boltzmann formalism~\cite{xiao2010berry}, and add a whole new dimension to the problem of electron transport in metals. It is important to point out that although a quantum linear-response formalism, such as the one developed by Kubo~\cite{kubo1957statistical} predicts the behavior in any system, the Boltzmann approach provides an intuitive and physically transparent picture. The study of electron transport in WSMs has explored the full potential of the topological Bloch-Boltzmann formalism, revealing surprising effects such as the chiral magnetic effect and chiral anomaly that were traditionally associated with high-energy physics. 
The reappearance of the same mathematical structures across different energy and length scales is an astounding feature uncovered in the Bloch-Boltzmann study of WSMs.

Initially, Weyl semimetals were predicted to only display negative longitudinal magnetoresistivity, which was considered to be a smoking gun signature of `chiral anomaly'. 
Several recent works have revisited this problem discovering that magnetoresistivity in LMC is more intricate, and crucially depends on the strength of internode scattering, the tilt of Weyl nodes, and also the presence of strain. Furthermore, it has been realized that anomalous Hall resistivity, planar Hall magnetoresistivity, Nernst, and thermoelectric conductivities can also provide signatures that reveal the underlying topology and chirality of the Bloch bands, making them striking phenomena of the manifestation of the quantum-mechanical nature of the Weyl wavefunction in macroscopic properties.
Several experimental works report nuanced behavior of magnetoresistivity in WSMs that is also fully consistent with the theoretical predictions reviewed in this work~\cite{flessa2023anisotropic,sonika2023chiral,tamanna2023hydrogen}.
In this review, we comprehensively discussed magnetotransport in WSMs, carefully examining the effects of different scattering mechanisms and strain.  Although the results presented are valid for a prototype model of only two nodes (or four nodes in case of inversion asymmetric WSMs), the physics is expected to hold for generic Weyl and Dirac systems. We end by stating that the development of toy models into comprehensive theories of real materials and their practical realizations remains a major challenge in contemporary condensed matter physics. 


\bibliography{biblio.bib}
\end{document}